\documentclass[aps,prd,onecolumn,groupedaddress,showpacs,nofootinbib,amssymb]{revtex4}
\usepackage{graphicx,bm,color}
\usepackage{wrapfig}
\usepackage{cancel}
\usepackage{here}
\usepackage{subfigure}
\usepackage{physics}
\usepackage{mathrsfs}
\usepackage{enumerate}
\usepackage{nccmath}
\usepackage{amsfonts}
\usepackage{amssymb}
\usepackage{mathtools}
\usepackage{comment}
\usepackage{CJK}
\usepackage[paperwidth=210mm,paperheight=297mm,centering,hmargin=2cm,vmargin=1.5cm]{geometry}

\newcommand{\nn}{\nonumber \\}
\newcommand{\e}{\mathrm{e}}
\newcommand{\E}{\mathrm{e}}
\newcommand{\h}{\tilde{h}}
\newcommand{\V}{\tilde{V}}
\newcommand{\Z}{\tilde{Z}}
\newcommand{\R}{\tilde{R}}

\newcommand{\Zm}{\mathcal{Z}}
\def\comb#1#2{{}_{#1} \mathrm{C} _{#2}}

\def\dress#1{| #1 \rangle \rangle}
\def\dressbra#1{\langle \langle #1 | }
\newcommand{\eps}{\varepsilon}
\newcommand{\Hil}{\mathscr{H}}
\newcommand{\ES}{\mathcal{S}}
\newcommand{\Fock}{\mathscr{H}_\mathrm{F}}
\newcommand{\floor}[1]{\lfloor #1 \rfloor}

\newcommand{\eqthesection}{\arabic{section}}
\makeatletter
 
  \@addtoreset{equation}{section}
 \makeatother

\allowdisplaybreaks[4]
\begin{document}
\begin{CJK}{UTF8}{}

\title{Dressed-Asymptotic States and QED Infrared Physics}
\author{Hideo~Furugori $^1$,\,\thanks{furugori.hideo@a.mbox.nagoya-u.ac.jp}
Shin'ichi~Nojiri$^{1,2}$\,\thanks{nojiri@gravity.phys.nagoya-u.ac.jp}}
\affiliation{$^{1)}$ Department of Physics, Nagoya University,
Nagoya 464-8602, Japan \\
$^{2)}$ Kobayashi-Maskawa Institute for the Origin of Particles
and the Universe, Nagoya University, Nagoya 464-8602, Japan}

\begin{abstract}

The dressed state formalisms, which incorporate interactions of soft particles into an asymptotic state, are known
as the prescriptions expected to solve the problem of infrared (IR) divergence in the quantum field theory (QFT).
A particularly famous example is the dressed state formalism proposed by Kulish and Faddeev in quantum electrodynamics (QED).
As pointed out by Hirai and Sugishita, however, this formalism has problems in gauge invariance and the IR divergence.
These problems are mainly caused by the existence of ghosts or unphysical photon modes.
Therefore, we start by studying the asymptotic states in the Coulomb gauge, which excludes ghosts and/or unphysical photon modes.
In this paper, we propose a formalism to construct the asymptotic states directly from the interaction of the theory by setting a sufficiently large time scale $T$.
In this dressed state formalism, we define the asymptotic interaction remaining at $\abs{t}>T$ in terms of some fixed order of $1/T$, and we are performing all calculations according to that order.
We study the asymptotic states in QED specifically, but we can formally apply the dressed state formalism proposed in this paper to any perturbative QFT.
We show that, at least in QED, we can construct divergence-free and unitary $S$-matrix using dressed states proposed in this paper.
Furthermore, we discuss the transition rate to show that we can predict experimental results.
We also show that the $\mathrm{U}(1)$ gauge symmetry of $S$-matrix leads to the QED large gauge symmetry,
and deviation of the expectation values of the vector potential between initial and final spacelike hypersurfaces emerges as a QED memory effect.
The dressed state formalism in this paper may give a unified and new insight into IR physics, including asymptotic symmetries, memory effects, and unitarity of the state evolution.

\end{abstract}

\maketitle
\end{CJK}

\section{Introduction \label{SEcI}}

The problem of the infrared (IR) divergence in the transition amplitudes is well-known in the gauge theories in four dimensions.
In quantum electrodynamics (QED), for example, this is a problem where loop corrections coming from low-energy photons (i.e., soft photons)
give an infinite phase to the transition amplitudes and make the transition probabilities going to be zero due to the IR divergences.
This problem has been solved by using the Bloch-Nordsieck (BN) formalism \cite{Bloch:1937pw}, where we assume that the physical final state
is a superposition of states with any number of soft photons because we cannot distinguish between a final state and final states adding an arbitrary number of soft photons to the final state.
In the BN formalism, the soft photon theorem, which connects amplitudes emitting soft photons with the original amplitude is essential
to recover the predictability of the theory \cite{Yennie:1961ad,Weinberg:1965nx}.

Interestingly, it has been discovered that this soft photon theorem is related to the asymptotic symmetry of the theory \cite{Strominger:2013lka,He:2014cra,Campiglia:2015qka,Kapec:2015ena}.
Adding to the memory effect, the relations between soft theorems and asymptotic symmetries and
memory effects are recently known as the ``IR triangle'', which is now actively investigated (for example, see \cite{Strominger:2017zoo}).
In the study, the asymptotic symmetry of QED called large gauge symmetry implies the existence of an infinite number of large charges with functional degrees of freedom.
By choosing the function of the large charge to a specific one and using the Ward-Takahashi identity, we can derive the soft photon theorem.
A similar relation between the Bondi-van der Burg-Metzner-Sachs (BMS) asymptotic symmetry \cite{Bondi:1960jsa,Sachs:1962wk} and linearized gravity theory has been found \cite{He:2014laa}.
These studies lead to the consideration that asymptotic symmetries shed new light on the information loss problem of the black hole (BH) \cite{Hawking:2016msc}.
Their idea is that an infinite number of soft charges could also exist on the BH, which generates soft photons and soft gravitons (and so on) in quantum theory,
and these soft particles may protect the unitarity of the time evolution from the formation to the evaporation of the BH.
In other words, the existence of the soft particles which save the IR divergence problem for transition processes in flat spacetime may play an important role even in curved spacetime.
We should note that, however, in the BN formalism, we cannot remove the IR divergence in the $S$-matrix although there is no divergence in the transition probability.
In other words, the unitarity of the quantum process is not guaranteed.
Therefore, it is appropriate that we use the dressed state formalisms
\cite{Chung:1965zza,Kibble:1968sfb,Kibble:1969ip,Kibble:1969ep,Kibble:1969kd,Kulish:1970ut,Ware:2013,Neuenfeld:2018,Hirai:2019gio}, by which we expect
to construct IR divergence-free $S$-matrix so that we can check the unitarity.

In the dressed state formalisms, we do not take asymptotic states as free particle states, but dressed states with an infinite number of soft particles.
A particularly famous example is the dressed state formalism by Kulish and Faddeev (KF) in QED \cite{Kulish:1970ut}.
Recalling that the soft theorem is derived from one of the large charges in the context of asymptotic symmetries,
we can see that the theorem is not equivalent to the asymptotic symmetry but included by the asymptotic symmetry.
Furthermore, there have been several studies that the dressed states are eigenstates of the large charges and naturally arise
by guaranteeing the conservation of the large charges \cite{Kapec:2017tkm,Choi:2017ylo}.
Hence, we can expect that dressed state formalisms may play an important role not only in the unitarity of the state evolution but also in the understanding of asymptotic symmetries.
On the other hand, as pointed out by Hirai and Sugishita \cite{Hirai:2019gio}, the dressed states given by Kulish and Faddeev have some problems in the gauge invariance and the IR divergence.
In other words, we should construct a more suitable dressed state formalism than the KF formalism.

Motivated by the above situation, we propose a dressed state formalism and we study the asymptotic states, $S$-matrix, transition rate, gauge symmetry, and memory effect in QED.
Contents of this paper are the following: In the next section, we briefly review the scattering problem in the quantum field theory (QFT).
Next, we review the KF formalism and point out some problems.
Especially we mention an important problem of the condition for the gauge invariance pointed out by Hirai and Sugishita \cite{Hirai:2019gio}.
In the Section \ref{SEcIII}, We propose a dressed state formalism to solve the problems and give the theory-independent definition of asymptotic states and $S$-matrix.
After that in Section \ref{SecIV}, we apply the dressed state formalism to QED as a specific example, and
we explicitly show that there is no divergence in the $S$-matrix proposed in this paper and give a physical transition rate.
We also discuss the gauge symmetry and memory effect.
The last section is devoted to the summary and prospects.

\section{Review on Kulish-Faddeev dressed state formalism and the problems \label{SecII}}

Although there are many papers on the dressed state formalism, the terminology and formulations used in the papers are not so unified.
In this section, we review the paper by Kulish and Faddeev \cite{Kulish:1970ut} in some detail and clarifies the terminology which we use to avoid confusion.

\subsection{Fock based $S$-matrix \label{SecIIA}}

First, we remind of the standard scattering problem in the QFT.
In the scattering problem, we predict the transition probability from an ``in'' state
$\ket{\Psi_{\alpha}^+}$ at the sufficiently far past\footnote{We set the reference time to the time origin $t=0$.}
to an ``out'' state $\ket*{\Psi_{\beta}^-}$ at the sufficiently
far future caused by the scatterings\footnote{
Here we use the Heisenberg picture.
}.
Here $\ket*{\Psi_{\alpha}^+}$ and $\ket*{\Psi_{\beta}^-}$ are eigenstates
of the Hamiltonian of the system with the energies $E_\alpha$ and $E_\beta$,
respectively\footnote{
We should note that, however, if these states are completely the eigenstates
of the Hamiltonian, there does not occur any scattering.
Hence, in rigorously,
we need to consider the transition from a wave packet
$\ket{\mathrm{in}}$ to another wave packet $\ket{\mathrm{out}}$, which are
defined by
\[
\ket{\mathrm{in}}\coloneqq \int d\alpha g_{\mathrm{\,in}}(\alpha)\ket{\Psi^+_\alpha}\, , \quad
\ket{\mathrm{out}}\coloneqq \int d\beta g_{\mathrm{\,out}}(\beta)\ket*{\Psi^-_\beta} \, .
\]
Here $g_{\mathrm{\,in}}(\alpha)$ and $g_{\mathrm{\,out}}(\beta)$ are amplitudes that do not vanish
and smoothly varying over some finite range $\Delta E$ of energies.
In this paper, however, we do not analyse by using the packets.
}.
We may decompose the Hamiltonian $H$ into the free Hamiltonian $H_0$ and
the interaction $V$ as $H = H_0 + V$.
It is difficult to follow the time-evolution
of states moment by moment in general if $V$ exists.
Then by assuming that the interaction is sufficiently weak in the far past and the far future,
we can use the particle picture and calculate the $S$-matrix which is defined by,
\begin{equation}
S_{\beta\alpha} \coloneqq \ip*{\Psi^-_\beta}{\Psi_\alpha^+}\, .
\label{defSmatrix}
\end{equation}
In the particle picture, the ``in'' and ``out'' states asymptotically
correspond to the free particle state $\ket{\Phi_\alpha}$ in a free theory.
By using the interaction picture\footnote{We use suffix ``I'' to express the
interaction picture.}, this correspondance can be seen as,
\begin{equation}
\lim_{t\to-\infty }\ket{\Psi^+_\alpha \left( t \right) }_\mathrm{I} \coloneqq
\lim_{t\to-\infty}\Omega \left( t \right) \ket{\Psi_{\alpha}^+} \simeq \ket{\Phi_\alpha} \, .
\label{particle picture}
\end{equation}
Here $\Omega \left( t \right) $ is a unitary operator connecting the fields in the Heisenberg picture
and the fields in the interaction picture, so that,
\begin{equation}
\mathcal{O}(\vec{x},t) = \Omega^\dagger(\vec{x},t) \mathcal{O}^\mathrm{I}(\vec{x},t) \Omega(\vec{x},t) \, .
\label{Omega}
\end{equation}
We can also express the asymptotic ``out'' state similarly and the $S$-matrix in the particle picture is given by
\begin{equation}
\ip*{\Psi^-_\beta}{\Psi_\alpha^+} \equiv
\mel*{\Psi^-_\beta}{\lim_{t^\prime\to\infty}\Omega^\dagger \left( t^\prime \right)\Omega \left( t^\prime \right)
\lim_{t\to-\infty}\Omega^\dagger \left( t \right) \Omega \left( t \right) }{\Psi_\alpha^+}
\overset{!}{\simeq} \bra{\Phi_\beta} \mathcal{S}_\mathrm{D} \ket{\Phi_\alpha}
\eqqcolon S_{\beta\alpha}^\mathrm{D} \, .
\label{Sininteraction}
\end{equation}
Here $\mathcal{S}_\mathrm{D}$ is the Dyson $S$-operator defined by
\begin{equation}
\label{SDoperator}
\mathcal{S}_\mathrm{D} \coloneqq \lim_{\substack{t^\prime\to\infty\\ t \to - \infty}}
\Omega \left( t^\prime \right)\Omega^\dagger \left( t \right)
= \mathcal{T}\exp \left[ -i\int_{-\infty}^\infty d\tau \,V^\mathrm{I} \left( \tau \right)  \right] \, .
\end{equation}
The notation of $\mathcal{T}$ expresses the time-ordered product.
The symbol ``$!$'' in Eq.~(\ref{Sininteraction}) denotes caution that we should take care of the validity.
Note that the last expression is derived by taking the limit $\abs{t}\to\infty$ before the unitary operator $\Omega(t)$ act on the states.
That is the starting point of our method, which we will see later.
We also define $S^\text{D}_{\beta\alpha}$ as the standard Fock based $S$-matrix calculated by using Dyson $S$-operator.
Eq.~(\ref{Sininteraction}) is the standard expression for the $S$-matrix in the particle picture.
Since free particle states are in the Fock space $\Hil_\text{F}$\,,\,$\ES_\mathrm{D}$ is the map from $\Hil_\text{F}$\,to $\Hil_\text{F}$.
The $S$-matrix $S^\text{D}_{\beta\alpha}$, sandwiched in the Fock states is not well-defined in QED due to the existence of the IR divergence.
The IR divergence problem in QED is that quantum corrections from soft photons attach an infinitely large phase factor
and an infinitely small coefficient to the $S$-matrix as in the Eq.~(\ref{soft corrected fock S matrix}).

\subsection{Effects of the asymptotic interaction and the Kulish-Faddeev $S$-operator
\label{SecIIB}}

The problem of an infinite phase appearing in the $S$-matrix was also known in the Coulomb scattering in non-relativistic quantum mechanics.
One way to solve this problem is known as the Dollard formalism \cite{Dollard:1964}, which redefines the $S$-matrix incorporating
the non-negligible influence of the Coulomb interaction, even at sufficiently far distances, as a cause of the problem.
By applying this formalism to relativistic field theory, Kulish and Faddeev have shown that the asymptotic state of QED
is not a free particle state but a dressed state clinging by countless photons and have used it to discuss the IR divergent-free $S$-matrix.
In other words, the dressed state formalism is a way to regard infrared divergence as a breakdown of the particle picture.

Now we will see the dressed state formalism by Kulish and Faddeev.
The QED Lagrangian they used is the Lorentz covariant form as follows.
\begin{equation}
\mathcal{L} = -\frac{1}{4}F_{\mu \nu}F^{\mu\nu} -\frac{1}{2\alpha} \left( \partial^\mu A_\mu \right)^2
 -\bar{\Psi} \left(\gamma^\mu \partial_\mu +m \right)\Psi
 -ie \bar{\Psi}\gamma^\mu\Psi A_\mu +\mathcal{L}_\mathrm{c}\,.
\label{bareLQ}
\end{equation}
Here, $\mathcal{L}_\mathrm{c}$ is counterterm for the renormalization.
Taking the Feynman gauge ($\alpha=1$) and moving to the canonical formalism,
we can find that vector field $\tilde{a}_\mu(x)\coloneqq \Omega^\dag \left( t \right) A_\mu(x)\Omega \left( t \right) $ and spinor field
$\psi(x)\coloneqq \Omega^\dag \left( t \right) \Psi(x)\Omega \left( t \right) $ with charge $e$
in the interaction picture are given by
\begin{align}
& \tilde{a}_\mu (x)= \int \frac{d^3 k}{ \left(2\pi \right) ^{3/2} \sqrt{2\omega}} \sum_{\tilde{h} }
\left[ \epsilon_\mu \left( \vec{k},\tilde{h} \right) \e^{ik\cdot x}a \left( \vec{k},\tilde{h} \right)
+ \epsilon_\mu ^\ast \left( \vec{k},\tilde{h} \right) \e^{-ik\cdot x}a^\dagger \left( \vec{k},\tilde{h} \right) \right]
\,,\, k^0= \abs*{\vec{k}}\eqqcolon \omega
\label{vector field} \\
& \psi(x) = \int\frac{d^3 p}{ \left(2\pi \right) ^{3/2}}\sum_{\sigma}
\left[ u_\sigma \left( \vec{p} \right) \E^{ip\cdot x}b_\sigma \left( \vec{p} \right)
+v_\sigma \left( \vec{p} \right) \E^{-ip\cdot x}d_\sigma^\dagger \left( \vec{p} \right) \right]
\,,\,p^0=\sqrt{m^2+\abs{\vec{p}}^2}\eqqcolon E_p\,.
 \label{spinorfield}
\end{align}
Here $h = \{+\,, -\}$ denotes transverse mode with helicity~$\pm1$, and $I=\{\text{S}\,,\text{L}\}$ denotes scalar mode and longitudal mode, respectively, and we define $\tilde{h}=\{h\,,I\}$.
Hereafter, we use the tilde ($\tilde{\,\,\,}$) to remind the reader that unphysical modes $I$ are included.
In (\ref{vector field}), $a\left( \vec{k},\h \right)$ is the annihilation operator of the $\h$ mode photon with momentum $\vec{k}$,
and $\epsilon_\mu \left( \vec{k},\h \right)$ is the polarization vector of the photon.
On the other hand, $b_\sigma \left( \vec{p} \right)$ and $d_\sigma \left( \vec{p} \right)$ are the annihilation operators of the
electron and the positron with spin $\sigma$ and momentum $\vec{p}$, respectively.
We put some notes in Appendix~\ref{App1} for the properties of the polarization vector $\epsilon_\mu \left( \vec{k},\h \right)$
and coefficient functions of spinor field $u_\sigma \left( \vec{p} \right),v_\sigma \left( \vec{p} \right)$.
In this theory, QED interacion is given by\footnote{We always use normal-ordered interaction.}
\begin{equation}
\tilde{V}^\mathrm{I} \left( t \right) =ie\int d^3x \tilde{a}_\mu(x)\bar{\psi}(x)\gamma^\mu\psi(x)\,.
\label{KFintQED}
\end{equation}
Kulish and Faddeev have considered how is the asymptotic interaction $\tilde{V}^\mathrm{I}_\mathrm{as}$, which we should take into account even in the far past or future.
By writing down the QED interaction explicitly by using the field expressions Eq.~(\ref{vector field}) and Eq.~(\ref{spinorfield}),
we get\footnote{We also use the notation that $a_\mu(k,\tilde{h})\coloneqq \epsilon_\mu \left( k,\tilde{h} \right) a \left(k,\tilde{h} \right)$.},
\begin{align}
\tilde{V}^\mathrm{I} \left( t \right) = ie & \int d^3 x \sum_{\h,\sigma,\sigma^\prime}
\int \frac{d^3k \,d^3p\, d^3q}{ \left(2\pi \right) ^{9/2}\sqrt{2\omega}}
\left[ a_\mu \left( \vec{k},\h \right)\e^{ik\cdot x} \right. \nn
& \times\left\{
\e^{i \left( p+q \right)\cdot x}\bar{v}_\sigma\left( \vec{p} \right)\gamma^\mu u_{\sigma^\prime} \left( \vec{q} \right)
d_\sigma\left( \vec{p} \right)b_{\sigma^\prime} \left( \vec{q} \right)
+ \e^{-i \left( p+q \right)\cdot x}\bar{u}_\sigma\left( \vec{p} \right)\gamma^\mu v_{\sigma^\prime}
\left( \vec{q} \right)b^\dagger_\sigma\left( \vec{p} \right)d^\dagger_{\sigma^\prime} \left( \vec{q} \right) \right.
\nn
& \quad \left. \left. + \e^{i \left( p-q \right)\cdot x}\bar{v}_\sigma\left( \vec{p} \right)\gamma^\mu v_{\sigma^\prime}
\left( \vec{q} \right)d^\dagger_\sigma\left( \vec{p} \right)d_{\sigma^\prime} \left( \vec{q} \right)
+ \e^{-i \left( p-q \right)\cdot x}\bar{u}_\sigma\left( \vec{p} \right)\gamma^\mu u_{\sigma^\prime}
\left( \vec{q} \right)b^\dagger_\sigma \left( \vec{p} \right)b_{\sigma^\prime} \left( \vec{q} \right)
\right\} + \left(\text{h.c.} \right) \right]\,.
\label{Vexplicit}
\end{align}
By integrating the above expression over $\vec{x}$, there appear the delta functions connecting the momenta.
Further integrating over the momentum $\vec{q}$, we find that, in the second line of Eq.~(\ref{Vexplicit}),
the coefficients of $t$ in the exponents are given by $\omega \pm \left( E_p+E_{p\pm k} \right)$.
We can neglect the second line in Eq.~(\ref{Vexplicit}) since the coefficients are always positive and violently oscillate in the far past or future.
On the other hand, we cannot neglect the corresponding coefficients in the third line in Eq.~(\ref{Vexplicit}) since
the coefficients of $t$ in the exponents are given by $\omega \pm \left( E_p-E_{p\pm k} \right)$, and these terms go to zero in the $\omega=0$ regions.
Kulish and Faddeev have defined the asymptotic interaction as a form where we set $\omega =0$, i.e.,
\begin{align}
\tilde{V}^\mathrm{I}_\mathrm{as}  \left( t \right)  \coloneqq e & \sum_{\tilde{h}} \int \frac{d^3 k}{ \left(2\pi \right) ^{3/2} \sqrt{2\omega}}\int d^3 p
\,\hat{\rho}\left( \vec{p} \right)\,v_p^\mu
\left[ a_\mu \left( \vec{k},\tilde{h} \right) \e^{ik\cdot v_p t}
+ \left( \text{h.c.} \right) \right]\, ,
\label{KFVas} \\
v^\mu_p\coloneqq & \frac{p^\mu}{E_p} \,,\quad
E_p \coloneqq \sqrt{m^2 + \abs{\vec{p}}^2}\,,\quad
\hat{\rho} \left( \vec{p} \right)\coloneqq \sum_\sigma \left[ b_\sigma^\dagger \left( \vec{p} \right)b_\sigma\left( \vec{p} \right)
 -d_\sigma^\dagger\left( \vec{p} \right)d_\sigma\left( \vec{p} \right) \right]\,,
\label{rho}
\end{align}
is the asymptotic interaction of QED which we should consider.
We note that the integration region of the photon momentum is unbounded\footnote{
This results in the ultraviolet (UV) divergence coming from the high momentum region.
Kulish and Faddeev have assumed that the divergence could be removed by the renormalization of mass.
However, there is room for discussion to deal with this UV divergence.
For example, there is a study regarding the detection limit of photons as the upper cutoff of the momentum value \cite{Carney:2017oxp}.
Or as we will propose later, we may regard the inverse of the time scale of the experiment as the upper cutoff.
}.
We can rewrite Eq.~(\ref{KFVas}) as
\begin{equation}
\tilde{V}^\mathrm{I}_\mathrm{as} \left( t \right)  = \int d^3 x\, \tilde{a}_\mu (x)j^\mu_\mathrm{cl}(x)\,,\quad
j^\mu_\mathrm{cl}(x) = e \int d^3 p \, v_p^\mu\hat{\rho}\left( \vec{p} \right)\delta^3 \left( \vec{x}-\vec{v_p}t \right) \, ,
\label{Vas density}
\end{equation}
and now we can easily find that photons couple with individual moving charged particles in the far past or far future.

Following the Dollard formalism, Kulish and Faddeev have investigated the asymptotic states by considering the asymptotic
time-evolution with asymptotic Hamiltonian $H_\mathrm{as}=H_0 + \V_\mathrm{as}$
in the Schr\"odinger picture\footnote{We use suffix ``S'' to express the Schr\"odinger picture.}.
Now, the operator $U_\mathrm{as} \left( t \right) $ which describes the time-development of the asymptotic
states satisfies,
\begin{equation}
i\frac{dU_\mathrm{as} \left( t \right) }{dt}=H^\mathrm{S}_\mathrm{as} \left( t \right) U_\mathrm{as} \left( t \right)  \, ,
\end{equation}
which can be solved by assuming $U_\mathrm{as} \left( t \right) =\e^{-iH^\mathrm{S}_0t}\Z \left( t \right) $.
Then $\Z \left( t \right) $ satisfies the
following equation,
\begin{equation}
 i\frac{d\Z \left( t \right) }{dt}=\V^\mathrm{I}_\mathrm{as} \left( t \right) \Z \left( t \right)  \, .
\end{equation}
The solution is given by
\begin{equation}
\Z \left( t \right) =\mathcal{T}\exp \left[ -i\int^t_0 d\tau\, \V^\mathrm{I}_\mathrm{as} \left( \tau \right)
\right] \, .
\label{Z}
\end{equation}
By using the expressions in Eq.~(\ref{KFVas}) and Eq.~(\ref{rho}), Eq.~(\ref{Z}) can be explicitly rewritten as
\begin{equation}
\tilde{Z} \left( t \right) =\exp \left[
 -i\int^t_0 d\tau\, \tilde{V}^\mathrm{I}_\mathrm{as} \left( \tau \right)
 -\frac{1}{2}\int^t_0 d\tau\int_0^\tau d\tau^\prime\,
\left[ \tilde{V}^\mathrm{I}_\mathrm{as} \left( \tau \right),\tilde{V}^\mathrm{I}_\mathrm{as} \left( \tau^\prime \right) \right]
\right]
 \eqqcolon\E^{\hat{\tilde{R}} \left( t \right) } \,\E^{i\hat{\theta} \left( t \right) }\,.
 \label{Zexplicit}
\end{equation}
The first term in the exponent is nowadays often called the dress operator $\hat{\R} \left( t \right) $:
\begin{equation}
\hat{\R} \left( t \right) \coloneqq
\int d^3k\int d^3 p \sum_{\h} \left[\hat{f} \left( p,k,\h;t \right) a^\dagger \left( \vec{k},\h \right) - \left( \text{h.c.} \right) \right]
\,,\quad \hat{\tilde{f}} \left( p,k,\h;t \right) \coloneqq \frac{e \hat{\rho}\left( \vec{p} \right)}{ \left(2\pi \right) ^{3/2} \sqrt{2\omega}}
\frac{p \cdot \epsilon^\ast \left( k,\h \right) }{k\cdot p}\e^{-ik \cdot v_p t} \, .
\label{dress op}
\end{equation}
Here, Kulish and Faddeev have removed a term $\hat{\R}(t=0)$ coming from the lower limit of the integration
in Eq.~(\ref{Zexplicit}), due to the consideration that conditions at $t =0$
should not affect asymptotic states.
For later use, we define,
\begin{equation}
D_{\hat{\tilde{f}} \left( t \right) } \coloneqq \exp\left[
\int d^3k\int d^3 p \sum_{\tilde{h}} \left[ \hat{\tilde{f}} \left( p,k,\tilde{h};t \right) a^\dagger \left( \vec{k},\tilde{h} \right)
 - \left(\text{h.c.} \right) \right]
\right]\,.
\label{def displacement}
\end{equation}
This operator is just a displacement operator for defining coherent states.
We put some notes in Appendix \ref{App2} for the properties of coherent states in non-relativistic quantum mechanics.
The second term in the exponent of Eq.~(\ref{Zexplicit}) is called the phase operator $i\hat{\theta} \left( t \right) $,
which gives an infinitely large phase and the explicit form is expressed by
\begin{equation}
i\hat{\theta} \left( t \right) \coloneqq i\frac{\e^2}{8\pi}\int d^3p \int d^3q\, \hat{\rho} \left( \vec{p} \right) \hat{\rho} \left(\vec{q} \right)
\frac{p\cdot q}{\sqrt{ \left( p\cdot q \right)-m^4}} \int^t_0 \frac{d\tau}{\tau} \, .
\label{phase op}
\end{equation}
Now, we know the explicit form of $U_\mathrm{as}$ and Kulish and Faddeev have defined the $S$-operator as follows,
\begin{align}
\label{Soperator}
\mathcal{S}_{\mathrm{KF}} \coloneqq&
\lim_{t^\prime\to\infty \,,\, t \to -\infty}
\mathcal{S}_{\mathrm{KF}}^{\mathrm{as}} \left( t^\prime,t \right)\, , \\
\mathcal{S}_{\mathrm{KF}}^{\mathrm{as}} \left( t^\prime,t \right)\coloneqq&
Z^\dagger \left( t^\prime \right)\mathcal{S} \left( t^\prime,t \right)Z \left( t \right)  \,,\quad
\mathcal{S} \left( t^\prime,t \right)\coloneqq
\mathcal{T}\exp \left[-i\int_{t}^{t^\prime}
d\tau \,V^\mathrm{I} \left( \tau \right)  \right]\, .
\end{align}
Since the dress operator $\hat{\tilde{R}} \left( t \right) $ goes to be zero when we take the limit $t\to \pm \infty$ because we have the relation
\begin{equation}
\lim_{t\to \pm \infty}\frac{1}{v_p\cdot k}\E^{ik\cdot v_p t}=\pm i\pi \delta \left( k\cdot v_p \right)=0\,,
\label{infdelta}
\end{equation}
we get
\begin{equation}
\mathcal{S}_\mathrm{KF} \overset{!}{=}
\E^{-i\hat{\theta}(\infty)}\,\mathcal{S}_\mathrm{D} \, \E^{i\hat{\theta}(-\infty)}\,,
\label{KF S operator}
\end{equation}
if we can take the limit $\abs{t}\to\infty$ in advance.

\subsection{Kulish-Faddeev asymptotic states and their $S$-matrix \label{SecIIC}}

Kulish and Faddeev have discussed the property of the Hilbert space $\Hil_\mathrm{as}$ where their asymptotic states lives in,
by considering that $\mathcal{S}_\mathrm{KF}^\mathrm{as}$ is the map from $\Hil_\mathrm{as}$ to $\Hil_\mathrm{as}$.
Assuming that $\mathcal{S} \left( t^\prime,t \right)$ is the map from $\Fock$ to $\Fock$,
we can regard $\Hil_\mathrm{as}$ as
\begin{equation}
 \Hil_\mathrm{as} = \tilde{Z}^\dag \left( t \right) \Fock \equiv D_{\hat{\tilde{f}} \left( t \right) }^\dag \Fock
 = D_{-\hat{\tilde{f}} \left( t \right) } \Fock=\E^{-\hat{\tilde{R}} \left( t \right) }\Fock\,,
 \label{asym Hil}
\end{equation}
because $\tilde{Z} \left( t \right) \tilde{Z}^\dag \left( t \right) =1$\footnote{
Now we consider the case that $\tilde{Z} \left( t \right)$ is not a well-defined operator on $\Fock$.
Indeed, we will see soon, $\tilde{Z} \left( t \right) $ is not a well-defined operator on $\Fock$ for states with charged particles.}.
Since the dress operator and the phase operator do not mix the charged particles and the photons,
it is useful that we decompose the Hilbert space into the Hilbert space of charged particles and
that of photons like $\Fock = \Hil_\psi \otimes \Hil_\gamma$.
Then we can express the vacuum state as a direct product of the vacuum state of the charged particle and the
vacuum state of the photon, such that $\ket{0}\eqqcolon\ket{0;\, \psi}\otimes \ket{0;\, \gamma}$.
To see the properties of the asymptotic states, we concretely consider the situation where the asymptotic state in the particle picture is given by
\begin{equation}
\ket{\Phi_\alpha}=
\ket{\psi_\alpha} \otimes \ket{0;\,\gamma }\coloneqq
b_{\sigma_1}^\dagger \left( \vec{p}_1 \right) \cdots b_{\sigma_N}^\dagger \left( \vec{p}_N \right)
d_{s_1}^\dagger \left( \vec{q}_1 \right) \cdots d_{s_{M}}^\dagger \left( \vec{q}_M \right)
\ket{0;\, \psi}\otimes \ket{0;\,\gamma} \, .
\label{in free particle}
\end{equation}
Now, we find
\begin{align}
D_{\hat{\tilde{f}} \left( t \right) }^\dag \ket*{\Phi_\alpha}
=& \ket{\psi_\alpha}\otimes\ket*{-\tilde{f}_\alpha \left( t \right) }\,,\quad
\ket*{-\tilde{f}_\alpha \left( t \right) }\coloneqq D^\dag_{\tilde{f}_\alpha} \left( t \right) \ket{0;\gamma} \, ,
\label{tildeindress}\\
D^\dag_{\tilde{f}_\alpha \left( t \right) } =&  D_{-\tilde{f}_\alpha \left( t \right) }\coloneqq \exp \left[
 - \sum_{\h}\int d^3k \left[ f_\alpha \left( k,\h;t )a^\dagger( \vec{k},\h \right)
 -\left( \text{h.c.} \right) \right] \right]\, ,
\label{tildein-coherent}\\
\tilde{f}_\alpha \left( k,\h;t \right) \coloneqq & \sum_{n\in \alpha}
\frac{e_n }{ \left(2\pi \right) ^{3/2} \sqrt{2\omega}}
\frac{p_n \cdot \epsilon^\ast \left( k,\h \right) }{k\cdot p_n}\E^{-ik \cdot v_n t}\,,\quad
v_n^\mu \coloneqq p^\mu_n /E_{n}\,,\quad
E_n \coloneqq \sqrt{m^2 +\abs{\vec{p}_n}^2}\,.
\label{deftildefalpha}
\end{align}
Here, $n$ is the label of the charged particle in the asymptotic state, and $e_n$ is the electric charge of the $n$-th particle.
We note that the asymptotic photon state $\ket*{-\tilde{f}_\alpha \left( t \right) }$ is just a coherent state.
We use the expression Eq.~(\ref{deftildefalpha}) when we add the species of charged particles by changing mass $m\to m_n$.
We should note, however, that we cannot apply the properties of coherent state straightforwardly when we add the hard photon
in the states because their asymptotic states contain photons with arbitrary momenta.
As in the quantum mechanical case, we can deform $ D^\dag_{\tilde{f}_\alpha \left( t \right) }$ to the normal-ordered form :
\begin{align}
D_{\tilde{f}_\alpha \left( t \right) }^\dag =& \exp\left[-\frac{1}{2}\sum_{\tilde{h}}
\int d^3 k \abs{\tilde{f}_\alpha \left( k,\tilde{h};t \right)}^2\right]\nn
&\times \exp\left[\sum_{\tilde{h}}\int d^3 k\tilde{f}_\alpha \left( k,\tilde{h};t \right) a^\dag \left( k,\tilde{h} \right) \right]
\exp\left[-\sum_{\tilde{h}}\int d^3 k\tilde{f}_\alpha^\ast \left( k,\tilde{h};t \right) a \left( k,\tilde{h} \right) \right]\,.
\label{normalordered displacement}
\end{align}
If there exists charged particles in the state of particle picture, $D_{\tilde{f}_\alpha \left( t \right) }^\dag$ is not a well-defined
operator on the Fock space because we find
\begin{equation}
\sum_{\tilde{h}}\int d^3 k \abs{\tilde{f}_\alpha \left( k,\tilde{h};t \right)}^2=\infty\,.
\label{infinite number of soft photons}
\end{equation}
Hence, $\Hil_\mathrm{as}\neq \Fock$ in general\footnote{If there is no charged particle in the state of particle picture, the asymptotic state is lived
in the Fock space since $\tilde{f}_\alpha=0$.
The Hilbert space $\Hil_\mathrm{as}$ called the von Neumann space is larger space than the Fock space.}.
Recalling that non-relativistic case, we can understand that the asymptotic states of QED are the dressed state with an infinite number of photons
because l.h.s. of Eq.~(\ref{infinite number of soft photons}) represents averaged number of photons.

Kulish and Faddeev have chosen the followng state $\dress{\Psi^\mathrm{KF}_\alpha}$ as a physically suitable asymptotic state which belongs to
$ \Hil_\mathrm{as}^\text{phys}\subset \Hil_\mathrm{as}$.
\begin{align}
\dress{\Psi_\alpha^\mathrm{KF}}\coloneqq &
D^\dagger_{\tilde{g}_\alpha} \ket*{\Phi_\alpha}\,,\quad
D_{\tilde{g}_\alpha} \coloneqq
\exp \left[\sum_{\tilde{h}}\int d^3 k \left[ g^\mu_\alpha \left( \vec{k} \right) \epsilon^\ast_\mu \left( \vec{k},\tilde{h} \right)  a^\dagger  \left( \vec{k},\tilde{h} \right)
 - \left(\text{h.c.} \right) \right] \right]\,,
\label{dressKF}\\
g^\mu_\alpha \left( \vec{k} \right)\coloneqq &
\sum_{n\in \alpha}\frac{\phi^n \left( k,p_n \right)e_n}{(2\pi ) ^{3/2}\sqrt{2\omega}}
\left( \,\frac{p_n^\mu}{p_n\cdot k}+c^\mu \right)\,.
\label{galpha}
\end{align}
Here,
$\phi^n \left( k,p_n \right)$ is an arbitrary function as long as the function satisfies the following convergence conditions:
\begin{align}
 -\frac{1}{2}\sum_{\tilde{h}}\int d^3 k\abs{\tilde{g}_\alpha \left( \vec{k},\tilde{h} \right)  -\tilde{f}_\alpha \left( k,\tilde{h};t \right)}^2\,<\infty
\,,\quad \Im\left[\sum_{\tilde{h}}\int d^3 k\tilde{g}_\alpha^\ast \left( \vec{k},\tilde{h} \right)  \tilde{f}_\alpha \left( k,\tilde{h};t \right)
\right]\,<\infty\,,
\label{convergence condition}
\end{align}
which ensure that $\dress{\Psi^\mathrm{KF}_\alpha}$ lives in $\Hil_\mathrm{as}$, are satisfied.
To satisfy these conditions, $\phi^n \left( k,p_n \right)$ should satisfies $\phi^n \left( k,p_n \right)=1$ in the neighborhood of $\omega = 0$.
We also use the expression $\tilde{g}_\alpha \left( k,\tilde{h} \right)\coloneqq g^\mu_\alpha \left( \vec{k} \right) \epsilon^\ast_\mu \left( \vec{k},\tilde{h} \right) $
and null vector $c^\mu$ satisfies $c\cdot k =-1$\footnote{We use the spacetime metric with signature $(-,+,+,+)$, and we can define
$c^\mu\coloneqq \frac{1}{2\omega} \left(1,-\hat{\vec{k}} \right) $.
Here, vectors with hat ($\hat{\,\,})$ denote unit vectors like $\hat{\vec{k}}=\vec{k}/\abs*{\vec{k}}$.}.
The null vector $c_\mu$ is introduced to dictate their asymptotic state $\dress{\Psi^\mathrm{KF}_\alpha}$ to satisfy the free Guputa-Bleuler (GB) condition :
\begin{equation}
k^\mu\epsilon_\mu \left( \vec{k},\tilde{h} \right)  a \left( \vec{k},\tilde{h} \right)  \dress{\Psi^\mathrm{KF}_\alpha}=0\quad \mbox{for any}\ \vec{k}\,,\,\tilde{h}
\label{GB}
\end{equation}
for excluding scalar mode and longitudinal mode.
The free GB condition leads to the condition $k_\mu g_\alpha^\mu \left( \vec{k} \right)=0$ for the $g_\alpha^\mu \left( \vec{k} \right)$ in Eq.~(\ref{dressKF}),
and we get $g_\alpha^\mu \left( \vec{k} \right)$ in Eq.~(\ref{galpha}).
Another feature is that Kullish and Faddeev formally have eliminated time dependence by using $\tilde{g}_\alpha$ instead of $\tilde{f}_\alpha ( t ) $.

Kulish and Faddeev have defined their $S$-matrix $S_{\beta\alpha}^{\mathrm{KF}}$ as
\begin{equation}
S_{\beta\alpha}^\mathrm{KF} =\lim_{t^\prime\to\infty , \, t\to -\infty}
\dressbra{\Psi^\mathrm{KF}_\beta}
\mathcal{S}_\mathrm{KF}^\mathrm{as} \left( t^\prime,t \right)
\dress{\Psi^\mathrm{KF}_\alpha}
=\lim_{t^\prime\to \infty , \, t\to -\infty} \bra*{\Phi_\beta}
D_{\tilde{g}_\alpha} \tilde{Z}^\dagger \left( t^\prime \right) \mathcal{S} \left( t^\prime,t \right) \tilde{Z} \left( t \right)
D^\dagger_{\tilde{g}_\alpha} \ket*{\Phi_\alpha}
\end{equation}
If we can take the limit $\abs{t}\to\infty$ in advance, $S^\mathrm{KF}_{\beta\alpha}$ is formally written by
\begin{equation}
S_{\beta\alpha}^\mathrm{KF} \overset{!}{=}
\dressbra{\Psi_\beta^\mathrm{KF}}
\E^{-i\hat{\theta}(\infty)}\,\mathcal{S}_\mathrm{D} \, \E^{i\hat{\theta}(-\infty)}
\dress{\Psi^\mathrm{KF}_\alpha}=
\bra*{\Phi_\beta}
\E^{-i\hat{\theta}(\infty)}D_{\tilde{g}_\alpha} \,\mathcal{S}_\mathrm{D} \,D^\dagger_{\tilde{g}_\alpha} \, \E^{i\hat{\theta}(-\infty)}
\ket*{\Phi_\alpha}\,.
\label{Naive SKF}
\end{equation}
As we will see later, infinite large phase factors in Eq.~(\ref{Naive SKF}) cancel with
the infinite large phase factor coming from the soft photon loop corrections.
This cancellation is what we can expect as in the Dollard formalism.
The KF $S$-matrix Eq.~(\ref{Naive SKF}) is accepted as an IR finite $S$-matrix, but their explanation is inadequate, as we will see in the next section.

\subsection{Some issues \label{SecIID}}

Chung have proved that the absolute value of the $S$-matrix is IR finite when we consider the asymptotic state is given by,
\begin{align}
\dress{\Psi_\alpha^\mathrm{Ch}}\coloneqq & \exp \left[\sum_{h=\pm}\int d^3 k \left[
f_\alpha \left(\vec{k},h;t=0 \right) a^\dag \left(\vec{k},h \right)- \left(\text{h.c.} \right) \right] \right]\ket{\Phi_\alpha}
=D_{f_\alpha(t=0)}\ket*{\Phi_\alpha}\,,
\label{Chung state}
\\
f_\alpha\left(\vec{k},h;t \right) \coloneqq & \sum_{n\in \alpha}
\frac{e_n }{ \left(2\pi \right) ^{3/2} \sqrt{2\omega}}
\frac{p_n \cdot \epsilon^\ast \left(k,h \right)  }{k\cdot p_n}\E^{-ik \cdot v_n t}\,.
\label{deffalpha}
\end{align}
In other words,
$\left| \left< \left< \left. \left. \Psi^\mathrm{Ch}_\beta \right| \mathcal{S}_\mathrm{D} \right|\Psi^\mathrm{Ch}_\alpha \right> \right> \right|$
is IR divergent free \cite{Chung:1965zza}.
Following this fact, Kulish and Faddeev have concluded that thier $S$-matrix Eq.~(\ref{Naive SKF}) is IR finite because they have considered
that their asymptotic state $\dress{\Psi^\mathrm{KF}_\alpha}$ of Eq.~(\ref{dressKF}) is equivalent to the Chung state $\dress{\Psi_\alpha^\mathrm{Ch}}$
of Eq.~(\ref{Chung state}) in the IR region.
As a matter of fact, however, $\dress{\Psi^\mathrm{KF}_\alpha}$ is not equivalent to $\dress{\Psi_\alpha^\mathrm{Ch}}$.
In the discussion of equivalence, Kulish and Faddeev decomposed $g_\alpha^\mu \left( \vec{k} \right)$ as
$g^\mu_\alpha \left( \vec{k} \right) =g_\alpha^+ \left( \vec{k} \right)\epsilon^\mu \left(\vec{k},+\right)+g_\alpha^- \left( \vec{k} \right)\epsilon^\mu \left(\vec{k},- \right)$,
and regarded $\dress{\Psi^\mathrm{KF}_\alpha}$ as a state without unphysical photon modes.
However, we should decompose
$g^\mu_\alpha \left( \vec{k} \right) =g_\alpha^+ \left( \vec{k} \right)\epsilon^\mu \left(\vec{k},+\right)+g_\alpha^- \left( \vec{k} \right)\epsilon^\mu \left(\vec{k},- \right)
+g_\alpha^0 \left( \vec{k} \right) k^\mu$, since adding the third term also holds the condition $g^\mu_\alpha k_\mu=0$.
In this case, we cannot eliminate the unphysical photon modes in $\dress{\Psi^\mathrm{KF}_\alpha}$.
In other words, the IR divergence of the KF $S$-matrix Eq.~(\ref{Naive SKF}) is non-trivial in their discussion due to the existence of the unphysical photon modes.

In addition, there is a problem with the free GB condition Eq.~(\ref{GB}) that Kulish and Faddeev have imposed on their asymptotic state.
Usually, the free GB condition is imposed on free particle states.
Is it really appropriate to impose the free GB condition on a state in the presence of asymptotic interactions?
Hirai and Sugishita have explored the asymptotic states in QED from the viewpoint of the BRS quantization with asymptotic interaction
and they have concluded that the gauge invariant condition that prohibits unphysical states in QED is different from the free GB condition
but the condition is given as follows \cite{Hirai:2019gio},
\begin{align}
&\lim_{\abs{t}\to\infty} \hat{G} \left(\vec{k},\tilde{h},t \right)  \dress{\Psi^\mathrm{as}} =0 \quad \mbox{for any $\vec{k}$ and $\tilde{h}$}\, ,
\label{BRS}
\\
&\hat{G} \left(\vec{k},\tilde{h},t \right) \coloneqq
k^\mu\epsilon_\mu a \left( \vec{k},\tilde{h} \right) -\hat{q} \left(\vec{k},t \right)\,,\quad
\hat{q} \left(\vec{k},t \right) \coloneqq
\int d^3p \frac{e\hat{\rho}\left( \vec{p} \right) \E^{-ik\cdot v_p t}}{ \left(2\pi \right) ^{3/2}\sqrt{2\omega}}\,.
\end{align}
The term $\hat{q}$ has been usually ignored under the assumption that interaction can be neglected in the far past or far future.
Although $\dress{\Psi^\mathrm{KF}_\alpha}$ does not satisfy this condition when there exsits any charged particle in the state,
the state
\begin{equation}
\dress{\Psi^\text{``KF''}_\alpha \left( t \right) } \coloneqq D_{\tilde{f}_\alpha \left( t \right) }\ket*{\Phi_\alpha}
\quad \mbox{up to phase}\,,
\label{so-called KFstate}
\end{equation}
which is often called as the KF state, satisfies the gauge invariant condition Eq.~(\ref{BRS})\footnote{
The asymptotic state $\dress{\Psi^\text{``KF''}_\alpha \left( t \right) }$ differs from the original KF asymptotic state $\dress{\Psi^\mathrm{KF}_\alpha}$
in the Hilbert space to which it belongs.
That is, $\dress{\Psi_\alpha^{\text{``KF''}}(t)}\notin\Hil_\mathrm{as}$ in general (cf. Eq.~(\ref{asym Hil})).}.
For instance, we can get $\dress{\Psi^\text{``KF''}_\alpha \left( t \right) }$ by omitting $c_\mu$ in $g_\alpha^\mu$ and by setting $\phi^n\left(k,p_n \right)=-\E^{-ik\cdot v_n t}$
though we come across time dependence of the state.
In that sense, Hirai and Sugishita have concluded that the free GB condition is not appropriate for
the gauge invariance condition of the physical asymptotic states, and we need not $c^\mu$.
Here, we note that the asymptotic state of Eq.~(\ref{BRS}) is a coherent state with unphysical photon modes.
Hence, we cannot apply Chung's arguments straightforwardly from the viewpoint of only in the ghost-free condition Eq.~(\ref{BRS}).
We also note that we cannot discuss the infinitely large phase factor messing the $S$-matrix only by using the ghost-free condition Eq.~(\ref{BRS})\footnote{
Later, based on the discussion of asymptotic symmetry, Hirai and Sugishita have suggested a candidate of the asymptotic state in QED, which is expected to remove
the IR divergence of the $S$-matrix \cite{Hirai:2020kzx}.}.

There are some other non-trivial points.
First of all, while Kulish and Faddeev have defined the asymptotic state as a form where we set $\omega =0$, the validity of applying this form to the entire photon energy region
needs to be discussed.
In this treatment, we need discuss also how to deal with the UV divergence from the asymptotic state.
In addition, when we derive the dressed operator from the asymptotic interaction, Kulish and Faddeev have removed the contribution from the lower end of the time integral by hand.
This treatment is also a bit arbitrary.
Also, the validity of taking the limit $\abs{t}\to \infty$ first and performing the calculation as in Eq.~(\ref{Naive SKF}) is questionable.
This is because, although the KF asymptotic state is formally time-independent, the convergence conditions Eq.~(\ref{convergence condition}) implicitly includes time.
If we take the limit $\abs{t}\to \infty$, then we get $\tilde{f}_\alpha ( t ) \to0$ from Eq.~(\ref{infdelta}).
So $\phi^n \left( k,p_n \right)$ must be zero to satisfy the convergence conditions.
In other words, the dressed state does not appear, and we are back to the $S$-matrix of the Fock basis Eq.~(\ref{Sininteraction}).
Another problem besides the construction method of the asymptotic state, the KF formalism does not have predictability due to the existence of the functional ambiguity
as $\phi^n \left( k,p_n \right)$\footnote{There is no way to determine the behavior of $\phi^n \left( k,p_n \right)$ in the far region from $\omega =0$ in the KF formalism.
This would undermine the predictability of the theory.}.

\section{Asymptotic states and $S$-matrix \label{SEcIII}}

Asymptotic states are dressed by unphysical photon modes when we consider manifestly covariant QED by adding a gauge-fixing term as in the papers by
Kulish and Faddeev or Hirai and Sugishita.
This situation leads the discussion of the IR divergent problem to be complicated.
The existence of ghosts and unphysical photon modes complicates the structure of the Hilbert space and the construction of asymptotic states
by generating debates about the existence of null vector $c_\mu$ and the existence of functional degrees of freedom like $\phi^n \left( k,p_n \right)$.
When approximation regarding an asymptotic state as a free particle state is good, we can naturally determine the form of corresponding
suitable asymptotic state from the free Hamiltonian of the theory\footnote{
For example, when we consider the scattering of the charged particles in the particle picture, it is often convenient to choose the asymptotic state as
$\ket*{\Phi_\alpha}= b_{\sigma_1}^\dagger \left( \vec{p}_1 \right) \cdots b_{\sigma_N}^\dagger \left( \vec{p}_N \right)
d_{s_1}^\dagger \left( \vec{q}_1 \right)\cdots d_{s_{M}}^\dagger \left( \vec{q}_M \right) \ket{0}$.}.
If we can directly determine the asymptotic state from the asymptotic Hamiltonian $H_\mathrm{as}$ of the theory without any discussion about the Hilbert space,
we may be able to construct the dressed state formalism with predictability without suffering from functional degrees of freedom.
In line with this idea, we discuss the gauge field in the Coulomb gauge, which naturally emerges from the representation theory of massless helicity one particle.
We have the following Lagrangian as a starting point.
\begin{equation}
\mathcal{L} = -\frac{1}{4}F_{\mu \nu}F^{\mu\nu} -\bar{\Psi} \left( \gamma^\mu \partial_\mu +m \right)\Psi
 -ie \bar{\Psi}\gamma^\mu\Psi A_\mu +\mathcal{L}_\mathrm{c}\,.
\label{bareL}
\end{equation}
Here, $\mathcal{L}_\mathrm{c}$ is counterterm of for the renormalization.
The gauge field $a_\mu (x)\coloneqq \Omega^\dag \left( t \right) A_\mu(x)\Omega \left( t \right) $ in the interaction picture is the physical photon field given by
\begin{align}
a_\mu (x)= \int \frac{d^3 k}{ \left(2\pi \right)^{3/2} \sqrt{2\omega}} \sum_{h }
\left[\epsilon_\mu \left(\vec{k},h \right)\e^{ik\cdot x}a \left(\vec{k},h \right)
+ \epsilon_\mu^\ast \left(\vec{k},h \right) \e^{-ik\cdot x}a^\dagger \left(\vec{k},h \right) \right] \, .
\label{photon field}
\end{align}
Properties of the polarization vector of the photon field are given in Appendix \ref{App1}.
In this case, we should consider the Dirac quantization, which simplifies the structure of the Hilbert space by prohibiting unphysical degrees of freedom,
though it complicates the discussion of phase space.
We will consider the gauge invariance of the theory after the discussion in the Coulomb gauge condition.

In the previous sections, we have seen that the existence of the asymptotic interaction is essential in the dressed state formalisms.
In the traditional calculation of the standard $S$-matrix of Eq.~(\ref{Sininteraction}), we use the standard
Fock states as a basis by taking $\abs{t} \to \infty$ before calculating.
However, in case there exists asymptotic interaction, we cannot replace the order of taking the limit.
We may now understand that this is the source of the IR divergence.
The KF formalism is an attempt to justify the replacement of the ordering of taking the limit by the new definition of the $S$-matrix based
on the dressed states containing an infinite number of photons.
As pointed out in Sec.~\ref{SecIID}, however, this justification is questionable due to the implicit time-dependence of the convergent conditions Eq.~(\ref{convergence condition})
and we could not be able to have dressed states in the limit of $\abs{t}\to\infty$.
In addition, there are some other points which seem unnatural, such as the fact that the KF formalism also requires arguments of the UV side
due to the existence of photons with any momenta, and the validity of the boundary condition $\hat{\R}(t=0)$ in the derivation of their dress operator.

In the following, we solve these problems by introducing time scale $T$ where
the ``in/out'' asymptotic states are defined.
Considering the possibility that we cannot neglect the interaction even at the far past/future $t=t_{\mathrm{I}/\mathrm{F}}$, we rewrite the ``in'' state and ``out'' state in equivalent forms:
\begin{equation}
\ket{\Psi_\alpha^+}\equiv\Omega^\dagger\left(t_I \right)\Omega\left(t_I \right)
\lim_{t \to -\infty}\Omega^\dagger(t)\Omega(t)\ket{\Psi_\alpha^+}\,,\quad
\ket*{\Psi_\beta^-}\equiv\Omega^\dagger\left(t_F \right)\Omega\left(t_F \right)
\lim_{t^\prime \to \infty}\Omega^\dagger(t^\prime)\Omega(t^\prime)\ket*{\Psi_\beta^-}
\label{equiv in out state}
\end{equation}
Since there is a time translation invariance in this stage, we can make $t_I\equiv -T\,,\,t_F\equiv T$.
Next, by using the correspondence (\ref{particle picture}), we obtain
\begin{equation}
\ket{\Psi_\alpha^+}\simeq \Omega^\dagger \left(t_I\right)\mathcal{S} \left(t_I,-\infty \right) \ket{\Phi_\alpha}\,,
\quad \mathcal{S} \left(t_I,-\infty \right)
=\lim_{t \to -\infty}
\mathcal{T}\exp \left[-i\int_{t}^{t_I} d\tau \,V^\mathrm{I} \left( \tau \right)  \right]\, .
\label{equiv particle picture}
\end{equation}
To investigate the interaction at $t=t_I$, we now substitute the expressions
of free fields into $V^\mathrm{I}\left(t_I \right)$ and examine the coefficients of $t_I$ in the exponents.
Terms with the coefficients of $\mathcal{O}(1)$ vanish due to the high-frequency oscillation or
the Riemann-Lebesgue lemma.
On the other hand, we may find that some coefficients are order
$\mathcal{O} \left(1/T \right)$ in some momenta regions.
Then, we expand the terms with such coefficients with $1/T$
in the above momenta regions, and we define the leading terms as the asymptotic interaction\footnote{
In the next section, we show the concrete formulation of the dressed state formalism in this paper by considering QED as an example.
}.
This point is different from the KF formalism where Kulish and Faddeev have used just a form of the interaction where we set $\omega =0$.
Hence the asymptotic interaction which remains at $t=t_I$ is given by
\begin{equation}
V^\mathrm{I} \left(t_I \right) = V^\mathrm{I}_\mathrm{as}  \left(t_I \right) + o \left( 1/T \right)\,.
\end{equation}
Since we have defined the asymptotic interaction as interaction terms at the leading order of $1/T$,
we should neglect higher-order terms in all calculations\footnote{
As we will mention in the Section \ref{SecIVB}, we can define the asymptotic interaction as terms up to (sub)$^k$-leading order of $1/T$ like
\[
V^\mathrm{I}\left(t_I \right) =V_\mathrm{as}^\mathrm{I}\left(t_I \right) + V_\mathrm{as,SL}^\mathrm{I}\left(t_I \right)+\cdots+ V_\text{as,S$^k$L}^\mathrm{I}\left(t_I \right)+o\left(1/T \right)\,,
\]
and this definition may make a difference if we leave $T$ as finite.
In this case, we should neglect (sub)$^{k+1}$-leading or higher-order terms in all calculations.
Throughout this paper, we adopt a framework considering only leading-order dressing.
}.
When $t\leq t_I$, $V^\mathrm{I} \left( t \right) $ is approximately given by $V_\mathrm{as}^\mathrm{I} \left( t \right) $
and the ``in'' state can be written by
\begin{equation}
\ket*{\Psi^+_\alpha} = \Omega^\dagger \left(t_I\right) \mathcal{Z} \left( t_I,-\infty \right)  \ket*{\Phi_\alpha}+o\left(1/T \right)
\,,\quad \mathcal{Z} \left( t_I,-\infty \right)  \coloneqq  \lim_{t \to -\infty} \mathcal{T}\exp \left[-i\int_{t}^{t_I}
d\tau \,V_\mathrm{as}^\mathrm{I} \left( \tau \right) \right]\,.
\label{def asymptotic evolution}
\end{equation}
Now we define the asymptotic state $\dress{\Psi_\alpha(-T)}$ of the ``in'' state $\ket*{\Psi^+_\alpha}$ at far past $t=t_I \equiv -T$
as follows
\begin{equation}
\dress{\Psi_\alpha(-T)} \coloneqq \Zm(-T,-\infty) \ket*{\Phi_\alpha}.
\label{def asymptotic states}
\end{equation}
We also define the asymptotic state $\dress{\Psi_\beta \left(T\right)}$ of the ``out'' state $\ket*{\Psi^-_\beta}$
at far future $t=t_F\equiv T$ same as asymptotic ``in'' state.
The $S$-matrix (\ref{defSmatrix}) coincides
with the limit of the following asymptotic $S$-matrix $S_{\beta\alpha}^{\mathrm{as}}(T) $
as we can see from the construction of the asymptotic state (\ref{equiv in out state}):
\begin{align}
&S_{\beta\alpha}^{\text{as}}(T)\coloneqq
\dressbra{\Psi_\beta(T)} \mathcal{S}\left(T,-T \right) \dress{\Psi_\alpha(-T)}
=\bra*{\Phi_\beta}\Zm^\dag(T,\infty)
\mathcal{S}\left(T,-T \right)
\Zm(-T,-\infty) \ket*{\Phi_\alpha}
\label{def asymptotic S-matrix}\\
&S_{\beta\alpha} \equiv \lim_{T\to \infty} S_{\beta\alpha}^{\text{as}}(T)
\label{S-matrix in our formalism}
\end{align}
If we leave $T$ as finite, the asymptotic $S$-matrix Eq.~(\ref{def asymptotic S-matrix}) is what we like to obtain.


\section{Example: Quantum Electrodynamics in Coulomb gauge\label{SecIV}}

In this section, we apply dressed state formalism in this paper to QED as an explicit example.

\subsection{Construction of the asymptotic states \label{SecIVA}}

As mentioned in the previous sections, we consider the QED in the Coulomb gauge condition to exclude unphysical states explicitly and discuss gauge invariance later.
The Coulomb gauge condition naturally appears from the representation theory of the Poincare group for the massless particle with helicity one particle.
Now by substituting the expressiions of the photon field Eq.~(\ref{photon field}) and the spinor field Eq.~(\ref{spinorfield}) into the interaction of the QED,
\begin{equation}
V^\mathrm{I} \left( t \right) =ie\int d^3x a_\mu(x)\bar{\psi}(x)\gamma^\mu\psi(x)\,,
\label{intQED}
\end{equation}
we obtain almost the same expression of Eq.~(\ref{Vexplicit}) except for there is no tilde.
Following the previous sections, we discuss the momentum integration region to derive the asymptotic interaction.
We claim that the coefficients of $t$ in the exponents in the third line of Eq.~(\ref{Vexplicit}) cannot be ignored because if $T$ is sufficiently large as $1/T \ll m$,
in the photon momentum region $\omega\le 1/T$, we find
\begin{equation}
\omega\pm\left(E_p-E_{p\pm k} \right)=
\omega \pm E_p \left[1- \left( 1\pm \frac{2\vec{k}\cdot \vec{p}+\vec{k}^2}{E_p^2} \right)^{1/2} \right]
=-k\cdot v_p +\mathcal{O} \left(\omega^2\sim \left(1/T \right)^2 \right)\,.
\label{omega approx}
\end{equation}
Assuming that the interaction only remains in the momentum region $\omega\le 1/T$,
the asymptotic interaction which is defined by the leading terms of the interaction expanded by $1/T$ as in Eq.~(\ref{omega approx}) is given by
\begin{equation}
V^\mathrm{I}_{\text{as}} (t_{I,F}) = e \sum_h \int_{\omega\le1/T} \frac{d^3 k}{(2\pi)^{3/2} \sqrt{2\omega}}\int d^3 p
\,v_p^\mu
\left[\epsilon_\mu \left(\vec{k},h\right)a \left(\vec{k},h\right) \e^{ik\cdot v_p t_{I,F}}
+ \left(\text{h.c.} \right)\right] \hat{\rho}(\vec{p})
\label{ourVas}
\end{equation}
This asymptotic interaction Eq.~(\ref{ourVas}) is different from the asymptotic interaction given by Kulish and Faddeev in Eq.~(\ref{KFVas}) because
in (\ref{ourVas}), there are no unphysical photon modes and there are only soft photons with energy $\omega \le 1/T$.
Hence there does not appear the UV divergence from asymptotic states in the formulation proposed in this paper.
Now, by following the arguments in Sec.~\ref{SecIIB}, we find that the asymptotic operator $\Zm \left(t_I,-\infty \right) $ for time-evolution is given by
\begin{align}
\Zm \left(t_I,-\infty \right)
= & \E^{\hat{R} \left(t_I\right)} \E^{i\hat{\theta} \left(t_I\right)},
\label{Omegaas decomposition}
\\
\hat{R} \left( t \right) \coloneqq&
\int_{\omega\le 1/T} d^3k\int d^3 p \sum_{h}\left[\hat{f} \left( p,k,h;t \right)
a^\dagger \left(\vec{k},h \right)- \left(\text{h.c.} \right)\right]
\,, \nn
\hat{f} \left( p,k,h;t \right) \coloneqq&  \frac{e \hat{\rho}\left( \vec{p} \right)}{ \left(2\pi \right) ^{3/2} \sqrt{2\omega}}
\frac{p \cdot \epsilon^\ast \left(k,h\right)  }{k\cdot p}\E^{-ik \cdot v_p t}\, ,
\label{our dress operator}
\\
i\hat{\theta} \left(t_I\right)\coloneqq&  \lim_{t\to -\infty}
i\frac{\e^2}{8\pi}\int d^3p \int d^3q\hat{\rho}\left( \vec{p} \right)\hat{\rho} \left( \vec{q} \right)
\frac{p\cdot q}{\sqrt{(p\cdot q)-m^4}}
\log\left(t_I/t\right)\,.
\label{our phase operator}
\end{align}
In contrast to the KF formulation which omits a term depending on the reference time $t=0$ by hand,
we get a dress operator (\ref{our dress operator}) which is naturally independent on any conditions
at $t=0$ by using the equation (\ref{infdelta}).
In addition, although the phase operator (\ref{our phase operator}) includes the logarithmic divergence,
the divergence can be controlled because the operator is finite before we take the limit $t\to-\infty$.

Now we construct the asymptotic state of the ``in'' state $\dress{\Psi_\alpha \left(t_I\right)}$.
When the asymptotic state in the particle picture includes only charged particles as in Eq.~(\ref{in free particle}), we obtain
\begin{align}
&\dress{\Psi_\alpha \left(t_I\right)}=  \lim_{t\to -\infty}\exp \left[ i\sum_{m,n\in\alpha}
\frac{e_me_n}{8\pi\beta_{mn}}\log\left(t_I/t\right) \right] \ket{\psi_\alpha}\otimes\ket{f_\alpha \left(t_I\right)}\, ,
\label{indress}\\
&\ket{f_\alpha \left(t_I\right)} = D_{f_\alpha \left(t_I\right)}\ket{0;\gamma}\,,\quad
D_{f_\alpha \left(t_I\right)}\coloneqq \exp \left[
\sum_h\int_{\omega \le 1/T} d^3k \left[f_\alpha \left(k,h;t_I \right) a^\dagger \left(\vec{k},h \right)
 - \left( \text{h.c.} \right) \right] \right]\,.
\label{in-coherent}
\end{align}
Here, $\beta_{mn}\coloneqq \sqrt{1-\frac{m^4}{\left( p_m\cdot p_n \right)^2}} $ is the relative velocity between the $n$-th and $m$-th particles.
We note that $f_\alpha  \left(k,h;t_I \right) $ is identical with Eq.~(\ref{deffalpha}) which has been proposed in the Chung states.
Since the asymptotic states include only soft photons in the formalism in this paper, we can straightforwardly extend the expressions
of the state Eq.~(\ref{indress}) to the cases including hard photons in the asymptotic states\footnote{
When we include the hard photon, by separating the energy region of the photon as
$\omega\le 1/T$ and $1/T<\omega$, we can define
$\ket*{0;\, \gamma} \eqqcolon \ket*{0;\, \gamma_{\omega\le 1/T}} \oplus
\ket*{0;\, \gamma_{\omega > 1/T}}$.
}.
Thus, $\ket{f_\alpha \left(t_I\right)}$ is just a coherent state of the {\it soft} photons.
That is, the asymptotic states in QED are dressed states clinging with an infinite number of soft photons.
Similarly, we can obtain $\dress{\Psi_\beta \left(t_F\right)}$ which is the asymptotic state of the ``out'' state.


\subsection{Asymptotic $S$-matrix and the physical transition rate \label{SecIVB}}

 From the definition of the asymptotic $S$-matrix Eq.~(\ref{def asymptotic S-matrix}), we can rewrite
\begin{equation}
S^\mathrm{as}_{\beta\alpha}(T)=\E^{i\theta_{\beta\alpha} \left(T,t \right)}
\bra{0}D^\dag_{\hat{f}(T)}
\prod_{m\in\beta} a \left(p_m,\sigma_m,\mathscr{N}_m \right)
\mathcal{S}\left(T,-T \right)
\prod_{n\in\alpha} a^\dag \left(p_n,\sigma_n,\mathscr{N}_n \right)
D_{\hat{f}\left(-T\right)}\ket{0}\,,
\label{asymptotic S-matrix}
\end{equation}
where $\mathscr{N}_n$ represents species of the $n$-th particle, i.e., photon, electron, and positron.
The phase factor $i\theta_{\beta\alpha} \left(T,t \right)$ coming from the dressed asymptotic states is given by
\begin{equation}
\theta_{\beta\alpha} \left(T,t \right)\coloneqq
\sum_{\substack{m,n\in \alpha\\ m,n\in \beta}} \frac{e_me_n}{8\pi\beta_{mn}}\log \left( T/t \right)\,.
\label{infinitephase}
\end{equation}
It diverges in the limit of $t\to \infty$.
Here, we sum up the case $m$ and $n$ are both in the initial state $\alpha$ or both in the final state $\beta$.
In (\ref{asymptotic S-matrix}), $\displaystyle \prod_{n\in m}a \left(p_m,\sigma_m,\mathscr{N}_m \right)$
denotes sequence of the annihilation operators in the final state $\beta$
and $\displaystyle \prod_{n\in\alpha} a^\dag \left(p_n,\sigma_n,\mathscr{N}_n \right) $ denotes sequence of the creation operators in the initial state $\alpha$.
For simplicity of the discussion, we consider the case that there are no hard photons in the asymptotic states.
We can, however, trivially add hard photons in the initial and final states because, in the formalism in this paper,
the creation and annihilation operators of hard photon commute with the dress operators.
Defining the symbol $\partial$ as any operator which satisfy $\bra{0}\partial =0$ and omitting arguments of momentum and spin, we find
\begin{align}
\begin{aligned}
&  D_{\hat{f}(T)}^\dag
\prod_{m^\prime, n^\prime}  b_{m^\prime}d_{n^\prime}
\mathcal{S}\left(T,-T \right)
\prod_{m,n}b^\dag_{m}d^\dag_{n}
D_{\hat{f}(-T)}\\
& =\prod_{m^\prime, n^\prime} b_{m^\prime}d_{n^\prime}
\left(
\mathcal{S}\left(T,-T \right) D^\dag_{f_\beta(T)}+ \left[D^\dag_{f_\beta(T)},\mathcal{S} \left( T,-T \right) \right]
\right)
D_{f_\alpha(-T)}  \prod_{m,n}b^\dag_{m}d^\dag_{n}\\
&= \prod_{m^\prime, n^\prime}  b_{m^\prime}d_{n^\prime}
\mathcal{S}\left(T,-T \right)\prod_{m,n}b^\dag_{m}d^\dag_{n}
D^\dag_{f_\beta(T)}D_{f_\alpha(-T)}
+ \prod_{m^\prime, n^\prime}  b_{m^\prime}d_{n^\prime}
\mathfrak{S}\left(T,-T \right)
\prod_{m,n}b^\dag_{m}d^\dag_{n}D^\dag_{f_\beta(T)}D_{f_\alpha(-T)} +\partial\,,
\end{aligned}
\label{proof of my S calculation}
\end{align}
where
\begin{align}
\begin{aligned}
&  \mathfrak{S}\left(T,-T \right)
\coloneqq \mathcal{T}\exp[e\int_{-T}^T d\tau \int d^3 x \mathfrak{A}_\mu (x)\bar{\psi}(x)\gamma^\mu \psi (x)]\,,\,\\
&\mathfrak{A}_\mu (x) \coloneqq
\sum_{n\in \beta}\int_{\omega \le 1/T }  \frac{e_nd^3 k }{(2\pi)^32\omega}
P_{\mu\nu}(\vec{k})\frac{p_n^\nu}{k\cdot p_n}\E^{-ik\cdot(x-v_nT)}\,.
\label{fraktul def}
\end{aligned}
\end{align}
In the third line, we have used the Eq.~(\ref{disentangled form of Displacement op}) and Eq.~(\ref{action of Ea Eadag}) in Appendix B.
When $T$ is sufficiently large, we can use approximation
\begin{equation}
T \frac{\sin \left[ \left(E_\beta-E_\alpha \right)T\right]}{\left(E_\beta-E_\alpha \right)T}\approx 2\pi\delta\left(E_\beta-E_\alpha \right)
\Rightarrow
\mel*{\Phi_\beta}{\mathcal{S}\left(T,-T \right)}{\Phi_\alpha}
\approx S_{\beta\alpha}^\mathrm{D}\,.
\label{Tapprox}
\end{equation}
Here, $S^\mathrm{D}_{\beta\alpha}$ is the standard Fock based $S$-matrix defined by Eq.~(\ref{Sininteraction}).
For the first term in Eq.~(\ref{proof of my S calculation}), we get
\begin{equation}
\prod_{m^\prime, n^\prime}  b_{m^\prime}d_{n^\prime}\mathcal{S}\left(T,-T \right)\prod_{m,n}b^\dag_{m}d^\dag_{n}D^\dag_{f_\beta(T)}D_{f_\alpha(-T)}
\approx (S^\mathrm{D}_{\beta\alpha} +\partial)D^\dag_{f_\beta(T)}D_{f_\alpha(-T)}
=S^\mathrm{D}_{\beta\alpha} D^\dag_{f_\beta(T)}D_{f_\alpha(-T)} +\partial\,.
\label{first term in Pf}
\end{equation}
For the second term in Eq.~(\ref{proof of my S calculation}), there is no photon operator and therefore, we cannot construct scattering diagram, i.e.
\begin{equation}
\prod_{m^\prime, n^\prime}  b_{m^\prime}d_{n^\prime}
\mathfrak{S}\left(T,-T \right)
\prod_{m,n}b^\dag_{m}d^\dag_{n}D^\dag_{f_\beta(T)}D_{f_\alpha(-T)}=\partial\,.
\label{second term of Pf}
\end{equation}
Combining Eq.~(\ref{first term in Pf}) and Eq.~(\ref{second term of Pf}) to the Eq.~(\ref{proof of my S calculation}), we get
\begin{equation}
S^\mathrm{as}_{\beta\alpha} (T) =\E^{i\theta_{\beta\alpha} \left(T,t \right)}
\ip*{f_\beta \left(T\right)}{f_\alpha \left(-T\right)}S^\mathrm{D}_{\beta\alpha}\,.
\label{Asymptotic S-matrix}
\end{equation}
This expression is consistent with Eq.~(64) in Chung' paper \cite{Chung:1965zza}.
Dividing the photon energy region of internal lines into the two regions, $\omega\in \left[ 1/t,1/T \right]$ and $\omega> 1/T$,
where $1/t$ is the IR cutoff, we obtain the following expression of $S^\mathrm{D}_{\beta\alpha}$ \cite{Weinberg:1995mt},
\begin{align}
S^\mathrm{D}_{\beta\alpha}=& \exp\left[
\frac{1}{2}\sum_{m,n}\frac{e_me_n\eta_m\eta_n}{ \left(2\pi \right) ^3}
\int_{1/t \le \omega \le 1/T} \frac{d^3 k}{2\omega} \alpha_{mn}(k)\right]
\exp\left[i\sum_{\substack{m,n\in \alpha\\m,n\in\beta}}
\frac{e_me_n}{8\pi\beta_{mn}}\log\frac{1/T}{1/t}\right]
S_{\beta\alpha}^\mathrm{D} \left(1/T\right)\,,\,\nn
& \alpha_{mn}(k) \coloneqq \frac{(p_m\cdot p_n)}{(p_m\cdot k)(p_n\cdot k)}\,.
\label{soft corrected fock S matrix}
\end{align}
In the first factor, the summation runs over all external lines and $\eta_n$ is the sign factor with the value
$+1$ for particles in the final state $\beta$ and $-1$ for particles in the initial state $\alpha$.
The third factor $S_{\beta\alpha}^\mathrm{D} \left(1/T\right)$ is the $S$-matrix only including the correction from
virtual photons with the energy $\omega > 1/T$ and this can be
finite by the prescription of the renormalization.
We note that the expression of Eq.~(\ref{soft corrected fock S matrix}) include only leading-order soft corrections.
The problem of the IR divergence is the problem that
the phase factor which appears in the second factor in (\ref{soft corrected fock S matrix}) diverges
and the first factor goes to be zero in removing the IR cutoff.

Finally, we calculate $\ip{f_\beta \left(T\right)}{f_\alpha \left(-T\right)}$, which is the inner product of the
coherent states and given by
\begin{align}
&\ip*{f_\beta(T)}{f_\alpha(-T)} = \Upsilon_{\beta\alpha}(T) \E^{i\chi_{\beta\alpha}(T)},
\label{dress correction}\\
&\Upsilon_{\beta\alpha}(T)\coloneqq
\exp\left[-\frac{1}{2}\sum_h\int_{1/t\le \omega \le 1/T} d^3k\abs{f_\beta\left(k,h;T \right)-f_\alpha\left(k,h;-T \right)}^2 \right],
\label{dapmingfactor}\\
&i\chi_{\beta\alpha}(T)\coloneqq i\Im\sum_h\int_{1/t\le \omega \le 1/T} d^3 kf_\alpha \left(k,h;-T \right)f^\ast_\beta\left(k,h;T \right).
\label{marginal phase}
\end{align}
Here we have used the IR cutoff $1/t$, again.
We begin with the calculation of $\Upsilon_{\beta\alpha} \left(T\right)$.
We can see
\begin{equation}
\label{FFF}
\sum_h\int d^3k \left| f_\beta \left(k,h;T \right) -f_\alpha \left(k,h;-T \right) \right|^2
=\sum_h\int d^3k\left(
\left| f_\alpha \left(-T\right) \right|^2+ \left| f_\beta \left(T\right) \right|^2-2\Re \left[ f_\alpha \left(-T\right)f_\beta^\ast \left(T\right) \right]
\right) \, .
\end{equation}
Now we have omitted $k$ and $h$ in the argument of $f$.
For the first term in the r.h.s. of (\ref{FFF}), we obtain
\begin{equation}
\sum_h\int d^3k\abs{f_\alpha \left(-T\right)}^2
=\frac{1}{\left(2\pi \right) ^3}\int \frac{d^3 k}{2\omega}
\left[
\sum_{m\in\alpha} e_m^2 \gamma_{mm}(k)
+ \sum_{\substack{m,n\in\alpha\\m\neq n}} e_me_n\gamma_{mn}(k)
\cos\left[ \left(v_m-v_n \right)\cdot kT\right] \right]\,,
\label{calcu1}
\end{equation}
\begin{equation}
\gamma_{mn}(k)
\coloneqq
\alpha_{mn}(k)
+\frac{p_m\cdot c}{p_m\cdot k}
+\frac{p_n\cdot c}{p_n\cdot k}\,.
\label{gammamn}
\end{equation}
We note that $c_\mu$ is identical with the null vector introduced by Kulish and Faddeev in (\ref{galpha}),
$c_\mu\coloneqq\frac{1}{2\omega} \left(1,-\hat{\vec{k}} \right)$.
For the third term in (\ref{FFF}), we also obtain
\begin{equation}
\sum_h\int d^3kf_\alpha \left(-T\right)f^\ast_\beta \left(T\right)
= \sum_{\substack{m\in\alpha \\n\in\beta}}\frac{e_me_n}{ \left(2\pi \right) ^3}\int
\frac{d^3k }{2\omega}\E^{i(v_m+v_n)\cdot kT} \gamma_{mn}(k)\,.
\label{calcu2}
\end{equation}
Using Eq.~(\ref{calcu1}) and Eq.~(\ref{calcu2}), we find
\begin{equation}
\hspace{-16.5pt}
\log\Upsilon_{\beta\alpha} \left(T\right)= -\int \frac{d^3k}{2\omega}
\left[ \sum_{m,n}\frac{\eta_m\eta_ne_me_n }{2 \left(2\pi \right)^3}\alpha_{mn}(k)
 -\sum_{m\neq n}\frac{\eta_m\eta_ne_me_n }{2 \left(2\pi \right)^3}\gamma_{mn}(k)
\left\{1-\cos \left[ \left( \eta_mv_m-\eta_nv_n \right) \cdot kT \right] \right\} \right].
 \label{value expression}
\end{equation}
Here, we use electric charge conservation $\displaystyle \sum_n\eta_ne_n=0$, which justified from symmetry of standard $S$-matrix $S_{\beta\alpha}^\mathrm{D}$.
The first term and the second term in Eq.~(\ref{value expression}) are $\order{\omega^{-1}}\sim \order{T}$ and $\order{\omega T}\sim \order{T^0}$ respectively.
Since the asymptotic states incorporate the asymptotic interaction defined by leading terms with $1/T$, we should take
\begin{equation}
\log\Upsilon_{\beta\alpha}(T)  \overset{\mathrm{L}}{=}
 -\int \frac{d^3k}{2\omega}
\left[ \sum_{m,n}\frac{\eta_m\eta_ne_me_n }{2(2\pi)^3}\alpha_{mn}(k)
\right]
\label{leading value expression}
\end{equation}
into the account as a correction from dressing.
Here, the symbol $\overset{\mathrm{L}}{=}$ denotes that the leading terms of l.h.s. equal to the leading terms of r.h.s.
We should take the second term in Eq.~(\ref{value expression}) into account when we consider the effect of the sub-leading asymptotic interaction and sub-leading soft radiative corrections.
Similarly, we find
\begin{equation}
i\chi_{\beta\alpha} \left(T\right)= \frac{i}{2 \left(2\pi \right) ^3}
\sum_{\substack{m\in\alpha,n\in\beta\\ m\in \beta,n\in \alpha}}
\int \frac{d^3 k}{2\omega}e_me_n\gamma_{mn}(k)\sin \left[ \left( v_m+v_n \right)\cdot kT \right]\,.
\label{phase expression}
\end{equation}
Phase correction from dressing is $i\theta_{\beta\alpha}(T,t) +i\chi_{\beta\alpha}(T)$ and since we can see in the $\omega$ integrand,
\begin{equation}
\theta_{\beta\alpha}(T,t)= \order{T},\,\quad \chi_{\beta\alpha}(T)=\order{\omega T} \sim \order{T^0}\,,\,
\label{phases order}
\end{equation}
phase factor $i\chi_{\beta\alpha}(T)$ is neglected in this framework.
Finally, leading-corrected asymptotic $S$-matrix $  S_{\beta\alpha}^{\mathrm{as,L}}(T)$ is given by
\begin{equation}
S_{\beta\alpha}^{\mathrm{as,L}}(T)=
S_{\beta\alpha}^\text{D}\left(1/T \right)\,.
\label{Sas caliculated}
\end{equation}
This expression is divergence-free.
Therefore, the asymptotic $S$-matrix Eq.~(\ref{Sas caliculated}) is well-defined as the operator which connects the asymptotic state at $t=-T$ to the asymptotic state at $t=T$.
If there is no state with energy $E<1/T$ in the Hilbert space for ``in'' and ``out'' states except for the vacuum state, we get
\begin{align}
\begin{aligned}
\int d\beta S_{\beta\gamma}^{\text{as},L^\ast}(T)S_{\beta\alpha}^{\mathrm{as,L}}(T)
&=
\int d\beta S_{\beta\gamma}^{\text{D}^\ast}\left(1/T \right)
S_{\beta\alpha}^{\text{D}}\left(1/T \right)
=\int _{E_\beta>1/T} d\beta \mel*{\Phi_\gamma}{\mathcal{S}_\mathrm{D}^\mathrm{hard}}{\Phi_\beta}
\mel*{\Phi_\beta}{\mathcal{S}_\mathrm{D}^\mathrm{hard}}{\Phi_\alpha}\\
&=\int_{E_\beta > 1/T} d\beta \ip*{\Psi^+_\gamma}{\Psi^-_\beta}\hspace{-5pt}\ip*{\Psi^-_\beta}{\Psi^+_\alpha}
=\ip*{\Psi_\gamma^+}{\Psi_\alpha^+}=\delta(\gamma-\alpha)\,.
\end{aligned}
\label{Unitarity}
\end{align}
That is,  the asymptotic $S$-matrix $  S_{\beta\alpha}^{\mathrm{as,L}}(T)$ is unitary.
Here, $\mathcal{S}_\mathrm{D}^\mathrm{hard}$ denotes Dyson $S$-operator including only particles with energy $E>1/T$.

At the end of this section, we discuss the transition rate.
We define $\Lambda_\text{D}$ as the detection limit energy of a photon i.e., we can detect a photon with $\omega>\Lambda_\text{D}$
and we cannot detect a photon with $\omega\le\Lambda_\text{D}$.
We also define $\Lambda$ as the resolution of energy measurement for the ``in'' and ``out'' states i.e.,
we can notice the soft photon emissions when the sum of soft photon energies excess $\Lambda$.
Now we have two possibilities,
\begin{enumerate}[(i)]
\item We cannot observe any soft photons i.e., $1/T\le\Lambda_\mathrm{D}$
\item We can observe an soft photon i.e., $1/T > \Lambda_\mathrm{D}$
\end{enumerate}
For the situation (i), we need to consider the cases that any number of soft photons with $\omega_i\in \left[ 1/T,\Lambda_\mathrm{D} \right]$ are emitted as long as $\sum_i \omega_i <\Lambda$.
Defining $\Gamma_{\beta\alpha}^\mathrm{D} \left(1/T\right)$ as the transition rate derived from $S^\mathrm{D}_{\beta\alpha} \left(1/T\right)$,
the physical transition rate $\Gamma_{\beta\alpha}^\text{as,L}(\Lambda_\mathrm{D},\Lambda)$ which we should consider to give predictions for the theory is given by
\begin{align}
&\Gamma_{\beta\alpha}^\text{as,L}(\Lambda_\mathrm{D},\Lambda)=\Gamma_{\beta\alpha}^\text{D}\left(1/T \right)
\sum_{N=0}^\infty \frac{(A_{\beta\alpha})^N}{N!}
\int_{1/T\le \omega_i\le \Lambda_\mathrm{D}\,,\,\sum_i \omega_i\le \Lambda}
\prod_{i=1}^N\frac{d\omega_i}{\omega_i}
=\mathcal{F}(\Lambda_\mathrm{D}/\Lambda;\,A_{\beta\alpha})(\Lambda_\mathrm{D} T)^{A_{\beta\alpha}}
\Gamma_{\beta\alpha}^\text{D}\left(1/T \right)\,,
\label{asymptotic physical rate }
\\
&A_{\beta\alpha}\coloneqq -\frac{1}{8\pi}\sum_{m,n}\frac{\eta_m\eta_ne_me_n}{\beta_{mn}}
\log(\frac{1+\beta_{mn}}{1-\beta_{mn}})>0\,,
\label{def beta alpha}
\\
&\mathcal{F}(x;A)\coloneqq \frac{1}{\pi}\int_{-\infty}^\infty
du \frac{\sin u}{u}\exp[A\int_0^x\frac{d\omega}{\omega}(\E^{i\omega u}-1)]
=1-\frac{A^2\Theta(x-1/2)}{2}\int_{1-x}^x\frac{d\omega}{\omega}\log(\frac{x}{1-\omega})+\cdots
\label{def F}
\end{align}
We note that $\Gamma^\mathrm{as,L}_{\beta\alpha} \left(\Lambda_\mathrm{D},\Lambda\right)$ is $T$-independent because
$\Gamma_{\beta\alpha}^\mathrm{D} \left(1/T\right)\propto  \left(1/T\right)^{A_{\beta\alpha}}$.
We can safely take the limit of $T\to\infty$ in this case.
This transition rate is exactly same with the transition rate in the conventional Bloch-Nordsieck formalism.

For the situation (ii), we cannot take the limit $T\to\infty$.
We assume that $T$ is fixed to be finite but sufficiently large value.
We can rewrite the asymptotic ``in'' state as
\begin{equation}
\dress{\Psi_\alpha(-T)}
= \lim_{t\to\infty}\exp[i\sum_{m,n\in\alpha}\frac{e_me_n}{8\pi\beta_{mn}}\log(T/t)]
\ket*{\psi_\alpha}\otimes[\ket*{f_\alpha^\mathrm{invisible}(-T)}\oplus \ket*{f_\alpha^\mathrm{visible}(-T)}]\,,
\label{detectable dress}
\end{equation}
where ``invisible'' and ``visible'' mean that we take only into account soft photons with energy $\omega\in [1/t,\Lambda_\mathrm{D}]$ and $\omega\in [\Lambda_\mathrm{D},1/T]$ respectively.
If we consider radiative correction from invisible photon with $\omega \in [1/t,\Lambda_\mathrm{D}]$ instead of it from soft photon with
$\omega\in[1/t,1/T]$ in the derivation of Eq.~(\ref{soft corrected fock S matrix}), we get
\begin{align}
S^\text{D}_{\beta\alpha}=&
\exp\left[
\frac{1}{2}\sum_{m,n}\frac{e_me_n\eta_m\eta_n}{(2\pi)^3}
\int_{1/t \le \omega \le \Lambda_\mathrm{D}} \frac{d^3 k}{2\omega}
\alpha_{mn}(k)\right]
\exp\left[i\sum_{\substack{m,n\in \alpha\\m,n\in\beta}}
\frac{e_me_n}{8\pi\beta_{mn}}\log\frac{\Lambda_\mathrm{D}}{1/t}\right]
S_{\beta\alpha}^\text{D}(\Lambda_\mathrm{D})\,,\,\\
S_{\beta\alpha}^{\mathrm{as,L}}(T)=&\E^{i\theta_{\beta\alpha}(1/\Lambda,t)+i\theta_{\beta\alpha}(T,\Lambda_\mathrm{D})}
\ip*{f^\text{invisible}_\beta(T)}{f_\alpha^\text{invisible}(-T)}\ip*{f^\text{visible}_\beta(T)}{f^\text{visible}_\alpha(-T)}S^\text{D}_{\beta\alpha}\nonumber \\
=& \exp\left[- \frac{1}{2}\sum_{m,n}\frac{e_me_n\eta_m\eta_n}{(2\pi)^3}
\int_{\Lambda_\mathrm{D} \le \omega \le 1/T} \frac{d^3 k}{2\omega}
\alpha_{mn}(k)\right]
\exp\left[i\sum_{\substack{m,n\in \alpha\\m,n\in\beta}}\frac{e_me_n}{8\pi\beta_{mn}}\log\Lambda_\mathrm{D}T\right]
S_{\beta\alpha}^\text{D}(\Lambda_\mathrm{D})\,.
\label{detectable Sas}
\end{align}
Since we also find
\begin{equation}
\sum_{m,n}\frac{e_me_n\eta_m\eta_n}{(2\pi)^3}
\int_{\Lambda_\mathrm{D} \le \omega \le 1/T} \frac{d^3 k}{2\omega}
\alpha_{mn}(k)= \sum_{m,n}\frac{e_me_n\eta_m\eta_n}{8\pi\beta_{mn}}\log(\frac{1+\beta_{mn}}{1-\beta_{mn}})\log\Lambda_\mathrm{D}T
=A_{\beta\alpha}\log\Lambda_\mathrm{D}T\,,
\label{detectable effect expression}
\end{equation}
the physical transition rate $\Gamma_{\beta\alpha}^\text{as,L}(T)$ which we should consider to give predictions for the theory is given by
\begin{equation}
\Gamma_{\beta\alpha}^\text{as,L}(T)=\left(\frac{1}{\Lambda_\mathrm{D}T}\right)^{A_{\beta\alpha}}\Gamma^\text{D}_{\beta\alpha}(\Lambda_\mathrm{D})\,.
\label{detectable phys rate}
\end{equation}
This transition rate is finite as long as $\Lambda_\mathrm{D} T$ is finite and $\Lambda_\mathrm{D}$-independent because $\Gamma_{\beta\alpha}^\text{D}(\Lambda_\mathrm{D})\propto (\Lambda_\mathrm{D})^{A_{\beta\alpha}}$.
If the detection limit of the photon detector is improved, we may distinguish the difference between the transition rate $\Gamma^\mathrm{as,L}_{\beta\alpha} \left(T\right)$ for the case (i) and
the transition rate $\Gamma_{\beta\alpha}^\text{D}(\Lambda_\mathrm{D})$ for the case (ii) in principle.

\subsection{Gauge invariance and large gauge symmetry \label{SecIVC}}

So far, we have argued about the asymptotic $S$-matrix in the Coulomb gauge.
In this section, we discuss gauge invariance.
Since $S_{\beta\alpha}^\mathrm{D}$ in Eq.~(\ref{Sas caliculated}) is nothing but the standard $S$-matrix and therefore gauge invariant,
we only need to discuss gauge invariance of $\ip*{f_\beta \left(T\right)}{f_\alpha \left(-T\right)}$.
For the coherent state of the soft photon $\ket{f_\alpha \left(\pm T \right)}$, if we consider the gauge
transformation $\epsilon_\mu \left(k,h\right)  \to \epsilon_\mu \left(k,h\right)  + \eps^{\ast}_\mp \left(k,h\right)  k_\mu$\footnote{
We use complex conjugate of $\eps$ just for making notation simple.}, we find
\begin{equation}
f_\alpha \left(k,h;\pm T \right)\to f_\alpha(k,h\pm T)+ \varepsilon^\mp_\alpha \left(k,h;\pm T \right) \, , \quad
\varepsilon_\alpha^\mp \left(k,h;\pm T \right)\coloneqq
\sum_{m\in\alpha}\frac{e_m}{ \left(2\pi \right) ^{3/2}\sqrt{2\omega}}\varepsilon_\mp \left(k,h\right)  \e^{\mp ik\cdot v_mT}
\label{gauge transformed f} \, ,
\end{equation}
and the inner product of the coherent states is transformed as,
\begin{align}
&\ip*{f_\beta(T)}{f_\alpha(-T)}\to \ip*{f_\beta(T)}{f_\alpha(-T)}\hspace{-2pt}\times\hspace{-4pt}
\ip*{\eps^-_\beta(T)}{\eps^+_\alpha(-T)}
\exp\left[
\sum_{h}\int_{1/t \le \omega \le 1/T} d^3 kZ_{\beta\alpha}\left(k,h;T \right)
\right],\\
&Z_{\beta\alpha}\left(k,h;T \right) \coloneqq
 - \Re\left[ \left( f^\ast_\beta(T)-f^\ast_\alpha(-T) \right)
\left(\eps_\beta^{-}(T)-\eps_\alpha^{+}(-T)  \right) \right]
+i\Im \left[ f_\alpha(-T)\eps_\beta^{-\ast }(T) +
f_\beta^\ast(T)\eps_\alpha^{+}(-T) \right].
\end{align}
We choose the function of the gauge transformation $\eps^\pm$ to be square integrable function in
three dimensional momentum space.
Hence, $\eps^\pm$ can be generally expanded by using the spherical harmonics $Y_{\ell m} \left(\hat{\vec{k}} \right)$ as follows,
\begin{equation}
\eps_\pm \left(k,h\right)  =\sum_{i=-1}^{\infty}
\sum_{m_\mathrm{i}=-\ell_\mathrm{i}}^{\ell_\mathrm{i}}\sum_{\ell_\mathrm{i}=0}^{\infty}
\omega^\mathrm{i}\chi^{\pm(i)}_{\ell_\mathrm{i}m_\mathrm{i}} \left(h\right)Y_{\ell_\mathrm{i}m_\mathrm{i}} \left(\hat{\vec{k}} \right) \, .
\label{general gauge func}
\end{equation}
Now we like to consider IR contribution of $\ip*{\eps^-_\beta(T)}{\eps^+_\alpha(-T)}$.
We will refer to gauge transformation Eq.~(\ref{general gauge func}) with $i=-1$,
\begin{equation}
\eps_a\left(k,h \right)=\omega^{-1}\xi^{a}\left(\hat{\vec{k}},h \right)
\,,\quad \xi^{a}\left(\hat{\vec{k}},h \right)
=\sum_{\ell, m}\chi^a_{\ell m}(h)Y_{\ell m}\left(\hat{\vec{k}} \right)
\neq 0\,,\quad
a= \begin{cases}
 - \text{ for ``out'' state} \\
+ \text{ for ``in'' state}
\end{cases}
\label{inout xi gaugeT}
\end{equation}
as the large gauge transformation in QFT.
Considering that large gauge transformation, we get
\begin{align}
\ip*{\eps_\beta(T)}{\eps_\alpha(-T)}
&\overset{\mathrm{L}}{=} \exp[
 -\frac{1}{2(2\pi)^3}\lim_{t\to\infty}\log\frac{1/T}{1/t}\sum_{m,n}\sum_h\int d\Omega_{\hat{\vec{k}}}
\left(
\eta_m\eta_ne_me_n \xi^{a_m}\left(\hat{\vec{k}},h \right)\xi^{a_n^\ast}\left(\hat{\vec{k}},h \right)
\right)
]\nn
&=\exp[-\frac{1}{2(2\pi)^3}\lim_{t\to\infty}\log\frac{1/T}{1/t}
\sum_h\sum_{\ell, m}\abs{Q_-\chi_{\ell m}^-(h)-Q_+\chi_{\ell m}^+(h)}^2]\,,
\label{large factor}
\end{align}
where $a_m$ denotes $-$ for $m\in \beta$ and $+$ for $m\in\alpha$ and $Q_+\,,\,Q_-$ is total electric charge of ``in'',``out'' state respectively.
Since the logarithm diverges in the limit $t\to \infty$, asymptotic $S$-matrix is nonzero only if
\begin{equation}
Q_-\chi^{-}_{\ell m} \left(h\right) =Q_+\chi^{+}_{\ell m} \left(h\right) \,\,
\mbox{for any }h, \ell, m
\label{asymptotic charge conservation}
\end{equation}
holds.
Conversely, if Eq.~(\ref{asymptotic charge conservation}) is true, $Z_{\beta\alpha}\left(k,h;T \right)$ is sub-leading with $1/T$ and hence,
we can conclude that asymptotic $S$-matrix is invariant under large gauge transformation.
Since $Q_-=Q_+$ due to the symmetry of the standard $S$-matrix $S^\mathrm{D}_{\beta\alpha}$,
the condition for gauge invariance in (\ref{asymptotic charge conservation}) requires
\begin{equation}
\xi^-\left(\hat{\vec{k}},h \right) = \xi^+\left(\hat{\vec{k}},h \right)
\label{large gauge symmetry}
\end{equation}
for the functions of the large gauge transformation for ``in/out'' asymptotic states.
These relations correspond to the existence of the large gauge charge in \cite{Kapec:2015ena}
and their large gauge transformation.

We discuss the asymptotic symmetry of QED in more detail.
We rewrite\footnote{
When we consider the Hilbert space, we only need to consider the rays (which identify the states different only from phase factors), so we ignored the phase.
}
the state of Eq.~(\ref{indress}) or counterpart for the ``out'' state as
$\dress{\Psi^{0}_\alpha \left(\pm T \right)}\equiv \ket*{\psi_\alpha}\otimes\ket*{f_\alpha^{0} \left(\pm T \right)}$
and we consider the large gauge transformation
\begin{equation}
\epsilon_\mu\left(k,h \right)\to\epsilon_\mu\left(k,h \right)+\omega^{-1}\xi\left(\hat{\vec{k}},h \right)k_\mu
\,,\,\xi\left(\hat{\vec{k}},h \right)\neq 0
\label{xi gaugeT}
\end{equation}
for $\dress{\Psi^{0}_\alpha \left(\pm T \right)}$ and obtain the asymptotic state written by
$\dress{\Psi^\xi_\alpha \left(\pm T \right)}\equiv\ket*{\psi_\alpha}\otimes \ket*{f_\alpha^\xi \left(\pm T \right)}$\,.
We can show
\begin{align}
\dressbra{\Psi^\eta_\beta(\pm T)}\dress{\Psi_\alpha^\xi(\pm T)}\overset{\mathrm{L}}{=}&
\begin{cases}
\delta(\beta-\alpha)&\text{for}\quad \eta\left(\hat{\vec{k}},h \right)=\xi\left(\hat{\vec{k}},h \right)\\
0& \text{for}\quad \eta\left(\hat{\vec{k}},h \right)\neq\xi\left(\hat{\vec{k}},h \right)
\end{cases}\,,
\label{ininoroutoutproduct}\\
\dressbra{\Psi^\eta_\beta(\pm T)}\dress{\Psi_\alpha^\xi(\mp T)}\overset{\mathrm{L}}{=}&
\begin{cases}
\ip*{f^\eta_\alpha(\pm T)}{f^\xi_\alpha(\mp T)}\delta(\beta-\alpha)=0&\text{for}\quad f_\alpha\neq 0 \,\,\text{or} \,\, f_\beta \neq 0\\
1& \text{for}\quad f_\alpha =f_\beta =0
\end{cases}
\,.
\label{inoutproduct}
\end{align}
Here, we have used the fact $\abs{\ip*{f_\alpha(T)}{f_\alpha(-T)}}=0$ unless $\alpha$ does not include any charged particle in the Eq.~(\ref{inoutproduct}).
We can observe that the Hilbert spaces $\mathscr{H}^+/\mathscr{H}^-$ which asymptotic ``in''/``out'' state lives in are divided
by the function $\xi \left(\hat{\vec{k}},h \right)$ into super-selection sectors.
That is, asymptotic states in QED belongs to
\begin{align}
& \mathscr{H}^+\equiv \oplus _\xi \mathscr{H}^+_\xi
\coloneqq \mathscr{H}_F\otimes \left(\oplus_\xi \mathscr{H}_\xi^\text{soft}(-T)\right)\,,\ \text{where} \
^\forall \alpha\,,\ \ket*{\Phi_\alpha}\in \mathscr{H}_F
\,,\ \ket*{f_\alpha^\xi(- T)}\in \mathscr{H}_\xi^\text{soft}(-T)\,,
\label{QED in Hilbert space}\\
& \mathscr{H}^-\equiv \oplus _\xi \mathscr{H}^-_\xi
\coloneqq  \mathscr{H}_F\otimes \left(\oplus_\xi \mathscr{H}_\xi^\text{soft}(T)\right)\,,\ \text{where} \
^\forall \alpha\,,\ \ket*{\Phi_\alpha}\in \mathscr{H}_F
\,,\ \ket*{f_\alpha^\xi( T)}\in \mathscr{H}_\xi^\text{soft}(T)\,.
\label{QED out Hilbert space}
\end{align}
Here, $\mathscr{H}_F$ is the Fock space which comprises charged particles and photons with $\omega>1/T$ (and vacuum).
We can also observe
\begin{equation}
\mathscr{H}^\mathrm{soft}_\xi(\pm T) \perp \mathscr{H}^\mathrm{soft}_\eta(\pm T)\ \text{if}\,\,\xi \neq \eta\, ,\quad
\mathscr{H}^\mathrm{soft}_\xi(\pm T) \perp \mathscr{H}^\mathrm{soft}_\eta(\mp T)\,,
\label{perpendicular}
\end{equation}
where symbol $\perp$ denotes that the inner product of any element of the l.h.s. and any element of the r.h.s. is zero.
The general asymptotic state containing the degrees of freedom of the large gauge function $\dress{\Psi^\xi_\alpha \left(\pm T \right)}$ belongs to $\mathscr{H}^\mp_\xi$, i.e.,
\begin{equation}
\dress{\Psi_\alpha^\xi(\pm T)}\in
\mathscr{H}^\mp_\xi =\mathscr{H}_\text{F}\otimes \mathscr{H}^\text{soft}_\xi(\pm T)\,.
\label{Hilbert space lived in asym state}
\end{equation}
Now we can understand that the operator $\mathcal{S}(T,-T)$ is the map from $\mathscr{H}^+_\xi$ to $\mathscr{H}^-_\xi$ and the asymptotic symmetry
in QED appears as the super-selection rule prohibiting the transition $\mathscr{H}^+_\xi\to\mathscr{H}^-_\eta$ if $\eta\neq \xi $.
This structure of the Hilbert spaces are quite different from the KF formalism where $\mathcal{S}(T,-T)$ is defined as the map from $\Fock$ to $\Fock$.

\subsection{QED memory effect in the dressed state formalism \label{SecIVCc}}

At the end of this section, we discuss QED memory effect in the dressed state formalism.
We consider the cace that there is no hard photons in the asymptotic states.
We can calculate the expectation value of the photon field at $t=T$ as
\begin{align}
\expval{a_\mu \left(\vec{x},T \right)}\coloneqq & \mel{f_\beta \left(T\right)}{a_\mu \left(\vec{x},T \right)}{f_\beta \left(T\right)} \nn
=& \int_{\omega\le1/T}\frac{d^3 k }{ \left(2\pi \right) ^32\omega}
\left[
\sum_{m\in\beta}\frac{e_m}{k\cdot p_m}p_m^\nu P_{\nu\mu} \left( \vec{k} \right)\E^{-\vec{k}\cdot \left( \vec{x}-\vec{v}_mT \right)}+ \left( \,\text{c.c.} \right)
\right] \nn
=& \sum_{m\in\beta}\frac{e_m}{ \left(2\pi \right) ^3}\int_{\omega\le1/T}\frac{d^3 k}{2\omega}
\left[
\frac{p_{m\mu}}{k\cdot p_m}+c_\mu +\frac{c\cdot p_m}{k\cdot p_m}k_\mu
\right]\cos \left\{ \vec{k}\cdot \left(\vec{x}-\vec{v}_mT \right) \right\} \, .
\label{vectorpotentialT}
\end{align}
Then we find that $\expval{a_0 \left(\vec{x},T \right)}=0$.
For simplicity of the discussion, we evaluate the value at $\vec{x}=0$ as\footnote{
In order to give predictions to actual experiment/observation, we should evaluate the values at two points like $\vec{x}=\pm \vec{x}_0$.
A more detailed and realistic analysis of the memory effect is beyond the scope of this paper.
}
\begin{align}
\expval*{\vec{a}(\vec{0},T)}=&
 -\sum_{m\in\beta}\frac{e_m \vec{v}_m}{8\pi^2}\int_0^{1/T}d\omega
\int_{-1}^1du \frac{\cos \left( \omega T v_m u \right)}{1-v_mu}
=-\sum_{m\in\beta}\frac{e_m \hat{\vec{v}}_m}{8\pi^2T}G \left(v_m \right)\,,\nn
G \left(v_m \right)\coloneqq & \left[\text{Ci} \left(1+v_m \right)-\text{Ci} \left(1-v_m \right)\right]\sin 1
 -\left[\text{Si} \left(1+v_m \right)-\text{Si} \left(1-v_m \right)\right]\cos 1
+2\text{Si} \left(v_m \right)\,,
\label{G-factor}
\end{align}
where $\text{Ci}(x)$ is cosine-integral function defined by
\begin{equation}
\mathrm{Ci}(x)\coloneqq -\int_x^\infty \frac{\cos t}{t}dt
=\gamma_\mathrm{E} +\log x +\mathrm{Cin}(x)\,,\quad
\mathrm{Cin}(x)\coloneqq \int^x_0dt\frac{1-\cos t}{t}\,,\quad
\gamma_\mathrm{E}: \text{Eular-Mascheroni constant}\,,
\label{Cosintegral}
\end{equation}
and $\text{Si}(x)$ is sine-integral function defined by
\begin{equation}
\text{Si}(x)\coloneqq \int_0^x \frac{\sin t}{t}dt\,.
\label{Sinintegral}
\end{equation}
The behavior of $G(v)$ is depicted in FIG.~\ref{G-factor 0to1} and FIG.~\ref{G-factor 0.9to1}.

\begin{minipage}{8cm}
\begin{figure}[H]
\begin{center}
\scalebox{0.7}{\includegraphics{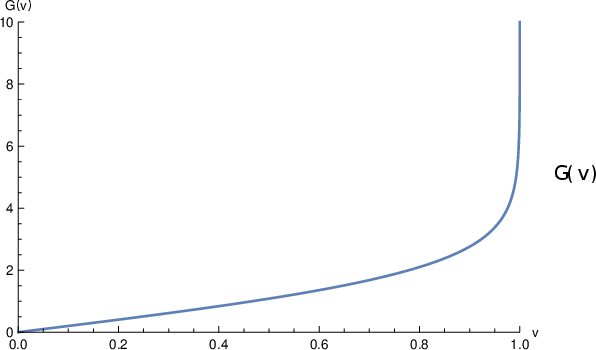}}
\caption{The behavior of $G(v)$ where $v\in[0,1]$}
\label{G-factor 0to1}
\end{center}
\end{figure}
\end{minipage}
\begin{minipage}{8cm}
\begin{figure}[H]
\begin{center}
\scalebox{0.7}{\includegraphics{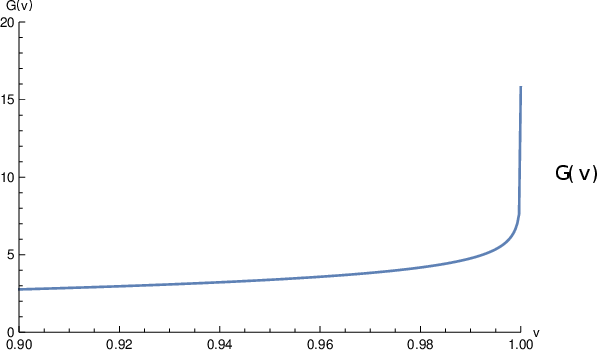}}
\caption{The behavior of $G(v)$ where $v\in[0.9,1]$}
\label{G-factor 0.9to1}
\end{center}
\end{figure}
\end{minipage}
\\
As shown in the figures, $G(v)$ diverges at $v = 1$, but even at $v=0.99$, we find $G(v) \sim \order{1}$.
Similarly we can find the expectation value of the vector potential at $x= \left( -T,\vec{0} \right)$ and the deviation is given by
\begin{equation}
\left< \delta \vec{a} \left( \vec{0} \right) \right>
= -\sum_{n}\frac{e_n \eta_n G \left(v_n \right)}{8\pi^2T}\hat{\vec{v}}_n\,.
\label{memory}
\end{equation}
We note that we cannot neglect this memory effect if we take $T$ as finite because this effect is leading with $1/T$ in the calculation for expectation values.
We may retrieve some information about the scattering process without observation of the scattering of charged particles by performing
AB effect-like experiments to detect the deviation of the vector potential.

We also comment the deviation of the electric field and the magnetic field.
Although there is a study which discusses the detection of the electromagnetic memory effect by using
the deviation of the electromagnetic field in classical theory \cite{Hamada:2018}, we find that no deviation appears
in both electric and magnetic fields in the formalism proposed in this paper.
For the electric part, this is obvious because we imposed the Coulomb gauge condition.
For the magnetic part, since $\vec{B}=\nabla \times \vec{a}$, we explicitly find
\begin{equation}
\left< \delta B \left(\vec{0} \right) \right>
= \sum_{m}\frac{e_m\eta_m}{2 \left(2\pi \right) ^3}\int _0^{1/T}d\omega
\omega \int d\Omega_{\vec{k}}
\frac{\vec{\hat{k}}\times \vec{v}_m}{1-\vec{\hat{k}}\cdot\vec{v}_m}
\sin \left( \omega T \vec{\hat{k}}\cdot \vec{v}_m \right)=0\,.
\label{delta B}
\end{equation}

Finally, we discuss the large gauge invariance of the QED memory.
We consider the gauge transformation of Eq.~(\ref{xi gaugeT}) and expand $\xi \left(\hat{\vec{k}},h \right)$ as follows,
\begin{equation}
\xi\left(\hat{\vec{k}},h \right)
=\sum_{\ell=0}^\infty\sum_{m=-\ell}^\ell\chi_{\ell m}(h)Y_{\ell m}\left(\hat{\vec{k}} \right)\,.
\label{xi expansion}
\end{equation}
Using associated Legendre polynomials $P^{m}_\ell(u)$
\begin{equation}
P^{m}_\ell(u) \coloneqq 2^{-\ell} \left( 1-u^2 \right)^{m/2}\sum_{j=0}^{\floor{\ell-m}}
\frac{(-1)^j (2\ell-2j)!}{j!(\ell-j)! (\ell-2j-m)!}u^{\ell-2j-m}\,,\,
\label{expression of associated Legendre polynomials}
\end{equation}
we can express spherical harmonics $Y_{\ell m}\left(\hat{\vec{k}} \right)$ as
\begin{equation}
Y_{\ell m}(\theta,\varphi)\coloneqq C_{\ell, m}s_m P_\ell^{\abs{m}}(\cos\theta)\E^{im\varphi}
\,,\,
C_{\ell, m}\coloneqq \sqrt{\frac{2\ell+1}{4\pi}\frac{(\ell-\abs{m})!}{(\ell+\abs{m})!}}\,,\,
s_m\coloneqq (-1)^{\frac{m+\abs{m}}{2}}\,.
\label{expression of spherical harmonics}
\end{equation}
Then, we can get the deviation from large gauge transformation $\left< \delta_\eps\vec{a} \left(\vec{x} \right) \right>$ as
\begin{align}
\hspace{-10pt}
\left< \delta_\eps\vec{a} \left(\vec{x} \right) \right>
= \sum_{n\in\alpha,\beta} \sum_h
\int_0^{1/T}\frac{e_n\eta_n d\omega}{ \left(2\pi \right) ^3 2\sqrt{2}}
\sum_{\ell, m}C_{\ell, m}s_m
\int d\Omega_{\vec{k}} P^{\abs{m}}_\ell(\cos\theta)
\left[ \chi_{\ell m} \left(h\right)\E^{i\psi_{mn}}\left(\vec{e}_\theta +ih \vec{e}_\varphi \right)
+\mathrm{\, (c.c.)}
\right]\,.
\label{delta a}
\end{align}
Here we have defined
\begin{equation}
\vec{\epsilon} \left(\vec{k},h \right)=\frac{1}{\sqrt{2}} \left( \vec{e}_\theta+ih\vec{e}_\varphi \right)\,,\quad
\vec{e}_\theta\coloneqq \mqty( \cos\theta\cos\varphi\\\cos\theta\sin\varphi\\-\sin\theta) \,,\quad
\vec{e}_\varphi\coloneqq \mqty(-\sin\varphi\\\cos\varphi\\0)\,,
\label{polarization expressions}
\end{equation}
\begin{equation}
\psi_{mn} \coloneqq m\varphi +\omega \delta x_n u\,,\quad
\delta x_n \coloneqq \abs{\vec{x}-\eta_n\vec{v}_nT}\,.
\label{def deltaxn}
\end{equation}
Using the relations
\begin{align}
\int_0^{2\pi}d \varphi \cos\varphi \cos(a+m\varphi)= \delta_{m\pm1} \pi \cos a \,, \quad &
\int_0^{2\pi}d \varphi \cos\varphi \sin(a+m\varphi)=\delta_{m\pm1} \pi \cos a \,,
\label{formula1} \\
\int_0^{2\pi}d \varphi \sin\varphi \cos(a+m\varphi)= -m\delta_{m\pm1} \pi \sin a \,, \quad &
\int_0^{2\pi}d \varphi \sin\varphi \sin(a+m\varphi)=m\delta_{m\pm1} \pi \sin a\,,
\label{formula2}  \\
\int_0^{2\pi} d \varphi \cos(a+m\varphi)= 2\delta_{m0}\pi \cos a \,,\quad &
\int_0^{2\pi} d\varphi \sin(a+m\varphi)= 2 \delta_{m0}\pi \sin a \,,
\label{formula3}
\end{align}
and defining
\begin{equation}
a_{\ell m}(h)\coloneqq \Re[\chi_{\ell m}(h)]\,\,,\,\,
b_{\ell m}(h)\coloneqq \Im[\chi_{\ell m}(h)]\,,
\end{equation}
we get
\begin{align}
&\left< \delta_\eps\vec{a} \left(\vec{x} \right) \right>
= \sum_{n\in\alpha,\beta} \sum_h
\int_0^{1/T}\frac{e_n\eta_n d\omega}{8\sqrt{2}\pi^2}
\sum_{\ell, m}C_{\ell, m}s_m
\int^1_{-1}du P^{\abs{m}}_\ell(u)
\biggl[\left(u \cos \psi_{0n}+h m\sin\psi_{0n} \right)
\left(a_{\ell m}(h)-b_{\ell m}(h)\right)\delta_{m \pm 1}  \vec{e}_1\nn
& -\left(h\cos\psi_{0n} +u m \sin \psi_{0n}\right)
\left(a_{\ell m}(h)+b_{\ell m}(h)\right)\delta_{m\pm1}\vec{e}_2
-2\sqrt{1-u^2}\left(a_{\ell0}(h)\cos\psi_{0n}-b_{\ell0}(h)\sin\psi_{0n}
\right )\delta_{m0}\vec{e}_3\biggl]\,,\,\nn
&\vec{e}_1\coloneqq(1,0,0)\,,\,\vec{e}_2\coloneqq (0,1,0)\,,\,\vec{e}_3\coloneqq (0,0,1)\,.
\label{expression of large gauge memory1}
\end{align}
Considering the fact that the parity of the associated Legendre polynomials is $P_\ell^m(-u)=(-1)^{m+\ell}P_\ell^m(u)$, finally we obtain
\begin{align}
\left< \delta_\eps\vec{a} \left(\vec{x} \right) \right>
&= \sum_{n\in\alpha,\beta} \sum_h
\int_0^{1/T}\frac{e_n\eta_n d\omega}{8\sqrt{2}\pi^2}
\int^1_{-1}du
\biggl[\,
\sum_{k=1}^\infty C_{2k,1}P^1_{2k}(u)\sum_{m=\pm1}s_m B_{2k,m}(h)
\left(u \cos \psi_{0n}+h m\sin\psi_{0n} \right)\vec{e}_1\nn
& - \sum_{k=1}^\infty C_{2k-1,1}P^1_{2k-1}(u)\sum_{m=\pm1}s_m A_{2k-1,m}(h)
\left(h\cos\psi_{0n} +u m \sin \psi_{0n}\right)\vec{e}_2\nn
&-2\sqrt{1-u^2}\sum_{k=0}^\infty \left(
C_{2k,0}P^0_{2k}(u)a_{2k,0}(h)\cos\psi_{0n}
 -C_{2k+1,0}P^0_{2k+1}(u)b_{2k+1,0}(h)\sin\psi_{0n}\right )\vec{e}_3
\biggl]\,,\,\nn
&
A_{\ell,m}(h)\coloneqq a_{\ell m}(h)+b_{\ell m}(h)\,,\,
B_{\ell,m}(h)\coloneqq a_{\ell m}(h)-b_{\ell m}(h)\,.
\label{large gauge transformed memory}
\end{align}
If QED memory is large gauge invariant physical quantity, the last expression Eq.~(\ref{large gauge transformed memory}) should be identically zero.
For the case that mode $\ell$ is even number, it leads to the conditions
\begin{align}
&\Re[\chi_{\ell0}(-h)]= -\Re[\chi_{\ell0}(h)]\,,\,\nn
&\Re[\chi_{\ell\pm1 }(-h)] -\Im[\chi_{\ell\pm1}(-h)]
= \Re[\chi_{\ell\mp1 }(h)] -\Im[\chi_{\ell\mp1}(h)]\,,\,
\label{even gauge inv condition}
\end{align}
and for the case that mode $\ell$ is odd number, it leads to the conditions
\begin{align}
&\Im[\chi_{\ell0}(-h)]= -\Im[\chi_{\ell0}(h)]\,,\,\nn
&\Re[\chi_{\ell\pm1 }(-h)] +\Im[\chi_{\ell\pm1}(-h)]
= -\Re[\chi_{\ell\mp1 }(h)] -\Im[\chi_{\ell\mp1}(h)]\,.
\label{odd gauge inv condition}
\end{align}
Therefore, if and only if $\abs{m} \le 1$ modes of the large gauge function $\xi \left(\hat{\vec{k}},h \right)$ satisfies the conditions
Eq.~(\ref{even gauge inv condition}) and Eq.~(\ref{odd gauge inv condition}),
QED memory is large gauge invariant physical quantity.
A discussion of the implications of extra conditions to preserve the large gauge symmetry of QED memory is a subject for future work.

\section{Summary and prospects \label{SecV}}

In this paper, by starting with the definition of the $S$-matrix,
we have proposed a dressed state formalism to construct the asymptotic state of ``in/out'' state at sufficiently far past/future at $t=\mp T$.
In this formalism, we define the asymptotic interaction remaining at $\abs{t}>T$ in terms of some fixed order of $1/T$, and incorporate it into the asymptotic states.
According to the order of $1/T$ in the asymptotic states, we determine the order of $1/T$, which we should take into account for all calculations.
We can construct the asymptotic states only from the interaction of the theory\footnote{
We have considered the interaction of QED $V^\mathrm{I}_\mathrm{QED} \left( t \right) $ as
\[
V^\mathrm{I}_\mathrm{QED} \left( t \right) =V^\mathrm{I} \left( t \right) \Theta \left( T-\abs{t} \right)+V_\mathrm{as}^\mathrm{I} \left( t \right) \Theta \left( \abs{t}-T \right)
\,,\quad \Theta(x)=\begin{cases}
1&\quad \text{for}\quad x\ge0\\
0&\quad \text{for}\quad x<0
\end{cases}\,.
\]
Of course there are the other way to smoothly connect $V^\mathrm{I} \left( t \right) $ to $V^\mathrm{I}_\mathrm{as} \left( t \right) $
without step function, however, the difference can affect at most only $\order{1/T}$ contribution.
On the other hand, as mentioned in a footnote of Sec.~\ref{SecIID}, the KF formalism has no predictability since we have no way to decide
non-IR behavior of functional ambiguity $\phi^n \left( k,p_n \right)$.
}.
We have also proposed the asymptotic $S$-matrix connecting the asymptotic state of ``in'' state at $t=-T$ with the asymptotic state of ``out'' state at $t=-T$.
We have shown that the asymptotic $S$-matrix is divergence-free and unitary at least in QED.
We have also discussed the transition rate and we have shown that the physical transition rate is exactly
the same as the transition rate obtained by the Bloch-Nordsieck formalism when we cannot observe soft photons with $\omega <1/T$.
In this case, we can safely take the limit of $T\to\infty$.
Also, we have seen that the dressed state formalism in this paper gives us a clear picture of how the relation known as the infrared triangle appears in QFT.
For example, we have shown that in QED,
\begin{enumerate}[(i)]
\item The large gauge transformation corresponds to the gauge transformation for the asymptotic states.
\item The large gauge symmetry implies that Hilbert space is divided by the large gauge functions into the super-selection sectors labeled by the functions.
\item The memory effect appears as the deviation of the expectation value of the vector potential caused by soft photons with information of charged particles.
\end{enumerate}
We may expect that such correspondence also appears in QFTs other than QED.
The fact that we can see such correspondences and the fact that we can construct a unitary $S$-matrix may show that
the formalism in this paper gives a unified and effective way to understand IR physics.

Furthermore, we have discovered that we can obtain a verifiable prediction in principle by leaving the sufficiently large time $T$ is finite.
Although we will need to more detailed discussion, we may verify the deviation of the transition rate if we have a photon detector
with a significantly high sensitivity and we may detect the memory effect in future experiments.

In addition to giving above perspective, the dressed state formalism in this paper is different from the KF formalism
in the following points.
\begin{enumerate}[(1)]
\item We do not suffer from the UV divergence from the asymptotic states.
\item Since the creation and annihilation operators of the soft photon commute with ones of the hard photon, we can add hard photons to the asymptotic states straightforwardly.
\item We do not need to require that the dress operator vanishes at $t=0$.
\item We do not have a functional ambiguity undermining predictability.
\end{enumerate}

We should note that the definitions of the asymptotic states and the asymptotic $S$-matrix proposed in this paper do
not depend, at least formally, on the kind of theories.
We may expect that the formulation in this paper may also give non-trivial results for the theories with massless particles besides QED.
For example, some studies indicate that asymptotic symmetry of scalar QED derives sub-leading
soft photon theorem \cite{Hirai:2018ijc,Campiglia:2019wxe}.
There are some arguments that there exist asymptotic symmetries besides
gauge theory \cite{Campiglia:2017dpg,Hamada:2017atr}.
Our formulation may give a new way of understanding the asymptotic symmetries
from the viewpoint of the asymptotic states in QFT.
We may also derive and give a new insight into the sub-leading (or higher) dressing which have proposed as counterpart of sub-leading soft theorem \cite{Choi:2019rlz}.
When we leave $T$ as finite, sub-leading terms may affect physical transition rate in case (ii) in Section \ref{SecIVB} and memory effect.
It is also interesting to apply the formulation in this paper to the linearized gravity theory.
In previous studies, the KF formalism or so-called KF state $\dress{\Psi^\text{``KF''}_\alpha \left( t \right) }$ of Eq.~(\ref{so-called KFstate})
are used to study linearized gravity theory \cite{Ware:2013,Choi:2017ylo}.
We may construct more suitable asymptotic states for analyzing the IR physics in the formalism in this paper.

Finally, we mention about the relation with the information paradox or information loss problem in the BH.
In the paper \cite{Hawking:2016msc}, it is shown that the photons with informations on the current
having fallen into the BH appear in the horizon and null infinity.
They claim that these photons are soft photons generated by the large gauge charges associated
with the asymptotic symmetry.
In this perspective, they discuss that the asymptotic symmetry and their charges are important
to understand the information paradox.
On the other hand, in the analyses in this paper, the asymptotic symmetry appears as just
a restriction on the gauge transformation for the soft photons in the asymptotic states,
that is, we cannot elicit the information of particles in the initial state from the asymptotic symmetry.
Instead, the asymptotic photon states themselves have the information via
the function of a soft photon coherent state $f_\alpha$.
 From this viewpoint, we may infer that when we consider the information paradox from the IR physics,
the charges are not so important but the dressed states are essentially important.
 From different contexts, there have been similar considerations by using the dressed states
\cite{Mirbabayi:2016axw,Gabai:2016kuf,Dominik:2018}.
Indeed, although we can obtain the soft photon theorem from the asymptotic symmetry,
the theorem itself cannot remove the IR divergence in the $S$-matrix. Therefore, we cannot discuss their unitarity.
As we have shown, the asymptotic $S$-matrix proposed in this paper is unitary and expected to be a starting point to consider the information paradox.
Also, there are some studies to investigate the differences generated by the dressed states
\cite{Carney:2017oxp,Carney:2018ygh}.
We expect the researches using the dressed states may become more important.

\section*{Acknowledgments}

This work is
supported by the JSPS Grant-in-Aid for Scientific Research (C) No. 18K03615
(S.N.).

\appendix

\renewcommand{\eqthesection}{\Alph{section}}

\section{Properties of the Coefficient Functions \label{App1}}
The gamma matrix is defined by
\begin{equation}
\gamma^\mu = (\gamma^0,\vec{\gamma})\,,\,
\gamma^0 \coloneqq -i\mqty(0 &I\\I &0)\,,\,
\vec{\gamma} \coloneqq -i\mqty(0 &\vec{\sigma}\\-\vec{\sigma} &0)\,,\,
\label{def of gamma}
\end{equation}
where $\sigma^i$'s are the Pauli matrices
\begin{equation}
\sigma^1\coloneqq \mqty(0&1\\1&0)\,,\quad
\sigma^2\coloneqq \mqty(0&-i\\i&0)\,,\quad
\sigma^3\coloneqq \mqty(1&0\\0&-1) \,.
\label{def of Pauli matrix}
\end{equation}
We also note that $\beta \coloneqq i\gamma^0$.
In the above definitions, the coefficient functions of spinor field satisfy the following equations,
\begin{align}
& \left( ip^\mu\gamma_\mu+m \right) u_\sigma\left( \vec{p} \right)=0\,,\quad
\bar{u}_\sigma\left( \vec{p} \right)u_{\sigma^\prime}\left( \vec{p} \right)
=\delta_{\sigma\sigma^\prime}\frac{m}{E_p}\,,\quad
\sum_{\sigma}\bar{u}_\sigma\left( \vec{p} \right)u_{\sigma}\left( \vec{p} \right)
=\frac{1}{2E_p} \left( -ip^\mu\gamma_\mu+m \right)\beta\,, \nn
& \left( ip^\mu\gamma_\mu-m \right)v \left( \vec{p},\sigma \right) =0\,,\quad
\bar{v}_\sigma\left( \vec{p} \right)v_{\sigma^\prime}\left( \vec{p} \right)
=-\delta_{\sigma\sigma^\prime}\frac{m}{E_p}\,,\quad
\sum_{\sigma}\bar{v}_\sigma\left( \vec{p} \right)v_\sigma\left( \vec{p} \right)
=-\frac{1}{2E_p} \left( ip^\mu\gamma_\mu+m \right)\beta\,.
\label{properties of spinor coefficients func}
\end{align}

We define the polarization vectors for the unphysical modes
$I=\{S,L\}$ as
\begin{equation}
\epsilon_\mu \left( \vec{k},S \right) \coloneqq \left( 1,\vec{0} \right)\,,\quad
\epsilon_\mu \left( \vec{k},L \right) \coloneqq \left( 0,\hat{\vec{k}} \right)\,.
\label{unphysical polarization}
\end{equation}
Then we find
\begin{equation}
\sum_{I}\epsilon_\mu^\ast \left( \vec{k},I \right) \epsilon_\nu \left( \vec{k},I \right)
=- \left( c_\mu k_\nu + k_\mu c_\nu \right)
\eqqcolon Q_{\mu\nu} \left( \vec{k} \right) \,,\quad
c^\mu\coloneqq \frac{1}{2\omega} \left( 1,-\hat{\vec{k}} \right)\,.
 \label{projection for onshell momentum}
\end{equation}
Since $Q^2=Q$ and $Q_{\mu\nu} \left( \vec{k} \right)k^\nu=k_\mu$, this operator is the projection operator to the on-shell momentum direction.
The properties of the physical modes $h=\{+,-\}$ are sufficient here if we know the following,
\begin{equation}
\epsilon_0 \left(\vec{k},h \right)=0\,,\quad
\vec{k}\cdot \vec{\epsilon}\, \left(\vec{k},h \right)=0\,,\quad
\sum_{h}\epsilon_\mu^\ast \left(\vec{k},h \right)\epsilon_\nu \left(\vec{k},h \right)
=\eta_{\mu\nu}-Q_{\mu\nu} \left( \vec{k} \right)
\eqqcolon P_{\mu\nu} \left( \vec{k} \right)\,.
\label{projection for transverse of onshell momentum}
\end{equation}
Since
$P^2=P$ and $P_{\mu\nu} \left( \vec{k} \right)k^\nu=0$, this operator is the projection operator to the transverse direction of the on-shell momentum.
We now obtain the following commutation relations
\begin{align}
\left[
a_\mu \left( \vec{k},\tilde{h} \right) ,a^\dag_\mu \left( \vec{k}^\prime,\tilde{h}^\prime \right)
\right]
=\eta_{\mu\nu}\delta_{\tilde{h}\tilde{h}^\prime}\delta^3 \left( \vec{k}-\vec{k}^\prime \right)  \,,\quad
\left[
a_\mu \left(\vec{k},h \right),a^\dag_\mu ( \vec{k}^\prime,h^\prime )
\right]
=P_{\mu\nu} \left( \vec{k} \right)\delta_{h h^\prime}\delta^3 \left( \vec{k}-\vec{k}^\prime \right)\,.
\label{commutation relations}
\end{align}
Since we can calculate the propagator as the vacuum expectation value of the time-ordered product of two fields at points $x$ and $y$,
we find that the propagators for the vector field $\tilde{a}_\mu$ and the photon field $a_\mu$ are given by
\begin{align}
 -i\tilde{\Delta}_{\mu\nu}(x-y)=& \frac{-i}{ \left(2\pi \right) ^4}
\int d^4k \frac{\eta_{\mu\nu}}{k^2-i\eps}\E^{i k\cdot (x-y)}\,,
\label{vector propagator}\\
 -i\Delta_{\mu\nu}(x-y)=& \frac{-i}{ \left(2\pi \right) ^4}
\int d^4k \frac{\tilde{P}_{\mu\nu}(k)}{k^2-i\eps}\E^{i k\cdot (x-y)}\,,
\label{photon propagator}
\end{align}
where $\tilde{P}_{\mu\nu}(k)$ is the projection operator extended to the off-shell momentum as\footnote{
We obtain this off-shell projection operator more explicitly when we rewrite Eq.~(\ref{projection for onshell momentum}) as
$Q_{\mu\nu} =\hat{\vec{k}}_\mu\hat{\vec{k}}_\nu-\hat{\vec{k}}_\mu t_\nu-t_\mu\hat{\vec{k}}_\nu$.
}
\begin{equation}
\tilde{P}_{\mu\nu}(k)= \eta_{\mu\nu}-\tilde{Q}_{\mu\nu}(k)\,,\quad
\tilde{Q}_{\mu\nu}(k)= \left(k_\mu k_\nu -k^0k_\mu t_\nu-k^0t_\mu k_\nu -k^2t_\mu t_\nu
\right)/\abs*{\vec{k}^{\,2}}\,,\quad
t^\mu=(1,0,0,0)\,.
\label{off-shell projections}
\end{equation}
Although the expression of Eq.~(\ref{photon propagator}) is not Lorentz covariant, we can use effective propagator as
Eq.~(\ref{vector propagator}) for calculations in the theory with the Lagrangian of Eq.~(\ref{bareLQ}).
Because the non-covariant term of the propagator can be eliminated by the non-covariant term in the Hamiltonian.

\section{Properties of the Coherent State in Quantum Mechanics \label{App2}}
Using the Baker-Campbell-Hausdorff formula
\begin{equation}
\E^A \E^B =\exp\left[A+B+\frac{1}{2}\left[ A,B \right]+\frac{1}{12} \left[A, \left[ A,B \right] \right]
 -\frac{1}{12} \left[B, \left[ A,B \right] \right] +\cdots \right]\,,
\label{BCH}
\end{equation}
we can reorder the displacement operator $D_\alpha$ as
\begin{equation}
D_\alpha\coloneqq \exp\left[\alpha a^\dag-\alpha^\ast a \right]
=\E^{-\frac{1}{2}\abs{\alpha}^2}\E^{\alpha a^\dag}\E^{-\alpha^\ast a}\,.
\label{disentangled form of Displacement op}
\end{equation}
The coherent state $\ket*{\alpha}$ with a parameter $\alpha \in\mathbb{C}$ is given by
\begin{equation}
\ket{\alpha}\coloneqq D_\alpha \ket{0}
=\E^{-\frac{1}{2}\abs{\alpha}^2}\E^{\alpha a}\ket{0} \, .
\end{equation}
Using the relation
\begin{equation}
\left[ a,f \left( a^\dag,a \right) \right]=\pdv{a^\dag} f \left( a^\dag,a \right)\,,
\label{a-derivative}
\end{equation}
we can easily find
\begin{equation}
a\ket{\alpha}=[a,D_\alpha]\ket{0}=\alpha\ket{0}\,,
\end{equation}
i.e., $\ket{\alpha}$ is the eigenstate of the annihilation operator with the eigenvalue $\alpha$.
With the particle number bases $\ket{n}$ given by
\begin{equation}
\ket{n}\coloneqq \frac{1}{\sqrt{n!}} \left( a^\dag \right)^n\ket{0}\,,
\label{n-number state}
\end{equation}
we can rewrite $\ket{\alpha}$ as
\begin{equation}
\ket{\alpha}=\E^{-\frac{1}{2}\abs{\alpha}^2}\sum_{n=0}^\infty \frac{\alpha^n}{\sqrt{n!}}\ket{n}\,.
\label{number based coherent state}
\end{equation}
Thus the coherent state is a superposition of any number of the particles.
The expectation value of the number of the particles is given by
\begin{equation}
\mel*{\alpha}{a^\dag a}{\alpha}=\abs{\alpha}^2\,.
\end{equation}
In the dressed states of QFT, the expectation value of the particle number is infinite due to the IR divergence of the counterpart of $\abs{\alpha}^2$.
It is frequently convenient to use the relations
\begin{equation}
\left[ \E^a,(a^\dag)^n \right]=\sum_{k=0}^{n-1}
\comb{n}{n-k}
\left(a^\dag\right)^k \E^a\,,\quad
\left[ a^n,\E^{a^\dag} \right]=\E^{a^\dag}\sum_{k=0}^{n-1}
\comb{n}{n-k}
a^k\,.
\label{action of Ea Eadag}
\end{equation}

Inner product of coherent states is given by
\begin{equation}
\bra*{\beta}\ket{\alpha}=\E^{-\frac{1}{2}(\abs*{\alpha}^2+\abs*{\beta}^2)}
\mel*{0}{\E^{\beta^\ast a}\E^{\alpha a^\dag}}{0}
=\E^{-\frac{1}{2}(\abs*{\alpha}^2+\abs*{\beta}^2)}
\sum_{n=0}^\infty \frac{(\alpha \beta^\ast)^n}{n!}\mel*{0}{\E^{\alpha a^\dag}}{0}
=\E^{-\frac{1}{2}(\abs*{\alpha}^2+\abs*{\beta}^2)}\E^{\alpha \beta^\ast}\,.
\label{inner product of coherent states-1}
\end{equation}
Or, equivalently
\begin{equation}
\bra*{\beta}\ket{\alpha}
=\E^{-\frac{1}{2}\abs{\beta-\alpha}^2}\E^{i\Im \alpha \beta^\ast}\,.
\label{inner product of coherent states-2}
\end{equation}
Thus coherent states satisfy the normality condition $\ip*{\alpha}{\alpha}=1$ but do not satisfy the orthogonality condition because $\ip*{\beta}{\alpha}\neq 0$ in general.
In QFT, however, the orthogonality often appears because the counterpart of $\abs{\beta-\alpha}^2$ can be divergent.
Inner product of parameter-translated coherent state is given by
\begin{align}
\begin{aligned}
\ip*{\beta+\eta}{\alpha+\xi}=\ip*{\beta}{\alpha}\hspace{-5pt}\ip*{\eta}{\xi}
\exp[-\Re[(\beta-\alpha)(\eta^\ast-\xi^\ast)]+i\Im[\alpha \eta^\ast+\xi \beta^\ast]\,]\,.
\label{inner product displaced}
\end{aligned}
\end{align}



\begin{thebibliography}{99}





\bibitem{Bloch:1937pw}
F.~Bloch and A.~Nordsieck,
Phys. Rev. \textbf{52} (1937), 54-59
doi:10.1103/PhysRev.52.54


\bibitem{Yennie:1961ad}
D.~R.~Yennie, S.~C.~Frautschi and H.~Suura,
Annals Phys. \textbf{13} (1961), 379-452
doi:10.1016/0003-4916(61)90151-8


\bibitem{Weinberg:1965nx}
S.~Weinberg,
Phys. Rev. \textbf{140} (1965), B516-B524
doi:10.1103/PhysRev.140.B516


\bibitem{Strominger:2013lka}
A.~Strominger,
JHEP \textbf{07} (2014), 151
doi:10.1007/JHEP07(2014)151
[arXiv:1308.0589 [hep-th]].


\bibitem{He:2014cra}
T.~He, P.~Mitra, A.~P.~Porfyriadis and A.~Strominger,
JHEP \textbf{10} (2014), 112
doi:10.1007/JHEP10(2014)112
[arXiv:1407.3789 [hep-th]].


\bibitem{Campiglia:2015qka}
M.~Campiglia and A.~Laddha,
JHEP \textbf{07} (2015), 115
doi:10.1007/JHEP07(2015)115
[arXiv:1505.05346 [hep-th]].


\bibitem{Kapec:2015ena}
D.~Kapec, M.~Pate and A.~Strominger,
Adv. Theor. Math. Phys. \textbf{21} (2017), 1769-1785
doi:10.4310/ATMP.2017.v21.n7.a7
[arXiv:1506.02906 [hep-th]].


\bibitem{Strominger:2017zoo}
A.~Strominger,
[arXiv:1703.05448 [hep-th]].


\bibitem{Bondi:1960jsa}
H.~Bondi,
Nature \textbf{186} (1960) no.4724, 535-535
doi:10.1038/186535a0


\bibitem{Sachs:1962wk}
R.~K.~Sachs,
Proc. Roy. Soc. Lond. A \textbf{A270} (1962), 103-126
doi:10.1098/rspa.1962.0206


\bibitem{He:2014laa}
T.~He, V.~Lysov, P.~Mitra and A.~Strominger,
JHEP \textbf{05} (2015), 151
doi:10.1007/JHEP05(2015)151
[arXiv:1401.7026 [hep-th]].


\bibitem{Hawking:2016msc}
S.~W.~Hawking, M.~J.~Perry and A.~Strominger,
Phys. Rev. Lett. \textbf{116} (2016) no.23, 231301
doi:10.1103/PhysRevLett.116.231301
[arXiv:1601.00921 [hep-th]].


\bibitem{Chung:1965zza}
V.~Chung,
Phys. Rev. \textbf{140} (1965), B1110-B1122
doi:10.1103/PhysRev.140.B1110


\bibitem{Kibble:1968sfb}
T.~W.~B.~Kibble,
J. Math. Phys. \textbf{9} (1968) no.2, 315-324
doi:10.1063/1.1664582


\bibitem{Kibble:1969ip}
T.~W.~B.~Kibble,
Phys. Rev. \textbf{173} (1968), 1527-1535
doi:10.1103/PhysRev.173.1527


\bibitem{Kibble:1969ep}
T.~W.~B.~Kibble,
Phys. Rev. \textbf{174} (1968), 1882-1901
doi:10.1103/PhysRev.174.1882


\bibitem{Kibble:1969kd}
T.~W.~B.~Kibble,
Phys. Rev. \textbf{175} (1968), 1624-1640
doi:10.1103/PhysRev.175.1624


\bibitem{Kulish:1970ut}
P.~P.~Kulish and L.~D.~Faddeev,
Theor. Math. Phys. \textbf{4} (1970), 745
doi:10.1007/BF01066485

\bibitem{Ware:2013}
J.~Ware, R.~Saotome, R.~Akhoury,
JHEP \textbf{10} (2013), 159
doi:10.1007/JHEP10(2013)159
[arXiv:1308.6285 [hep-th]].

\bibitem{Neuenfeld:2018}
D.~Neuenfeld
[arXiv:1810.11477 [hep-th]].

\bibitem{Hirai:2019gio}
H.~Hirai and S.~Sugishita,
JHEP \textbf{06}, 023 (2019)
doi:10.1007/JHEP06(2019)023
[arXiv:1901.09935 [hep-th]].


\bibitem{Kapec:2017tkm}
D.~Kapec, M.~Perry, A.~M.~Raclariu and A.~Strominger,
Phys. Rev. D \textbf{96} (2017) no.8, 085002
doi:10.1103/PhysRevD.96.085002
[arXiv:1705.04311 [hep-th]].


\bibitem{Choi:2017ylo}
S.~Choi and R.~Akhoury,
JHEP \textbf{02} (2018), 171
doi:10.1007/JHEP02(2018)171
[arXiv:1712.04551 [hep-th]].


\bibitem{Dollard:1964}
J.~D.~Dollard,
Journal of Mathematical Physics \textbf{5} (1964) 729-738
doi:10.1063/1.1704171


\bibitem{Carney:2017oxp}
D.~Carney, L.~Chaurette, D.~Neuenfeld and G.~W.~Semenoff,
Phys. Rev. D \textbf{97} (2018) no.2, 025007
doi:10.1103/PhysRevD.97.025007
[arXiv:1710.02531 [hep-th]].

\bibitem{Hirai:2020kzx}
H.~Hirai, S.~Sugishita,
JHEP \textbf{02} (2020), 02531
doi:10.1007/JHEP02(2021)025
[arXiv:2009.11716 [hep-th]].


\bibitem{Weinberg:1995mt}
S.~Weinberg,
``The Quantum theory of fields. Vol. 1: Foundations,''


\bibitem{Hamada:2018}
Y.~Hamada and S.~Sugishita,
JHEP \textbf{07} (2018), 017
doi:10.1007/JHEP07(2018)017
[arXiv:1803.00738 [hep-th]].

\bibitem{Hirai:2018ijc}
H.~Hirai and S.~Sugishita,
JHEP \textbf{07} (2018), 122
doi:10.1007/JHEP07(2018)122
[arXiv:1805.05651 [hep-th]].

\bibitem{Campiglia:2019wxe}
M.~Campiglia and A.~Laddha,
JHEP \textbf{10} (2019), 287
doi:10.1007/JHEP10(2019)287
[arXiv:1903.09133 [hep-th]].


\bibitem{Campiglia:2017dpg}
M.~Campiglia, L.~Coito and S.~Mizera,
Phys. Rev. D \textbf{97} (2018) no.4, 046002
doi:10.1103/PhysRevD.97.046002
[arXiv:1703.07885 [hep-th]].


\bibitem{Hamada:2017atr}
Y.~Hamada and S.~Sugishita,
JHEP \textbf{11} (2017), 203
doi:10.1007/JHEP11(2017)203
[arXiv:1709.05018 [hep-th]].

\bibitem{Choi:2019rlz}
R.~Choi and A.~Akhoury,
  JHEP \textbf{09} (2019), 031
  doi:10.1007/JHEP09(2019)031
  [arXiv:1907.05438 [hep-th]]


\bibitem{Mirbabayi:2016axw}
M.~Mirbabayi and M.~Porrati,
Phys. Rev. Lett. \textbf{117} (2016) no.21, 211301
doi:10.1103/PhysRevLett.117.211301
[arXiv:1607.03120 [hep-th]].


\bibitem{Gabai:2016kuf}
B.~Gabai and A.~Sever,
JHEP \textbf{12} (2016), 095
doi:10.1007/JHEP12(2016)095
[arXiv:1607.08599 [hep-th]].


\bibitem{Dominik:2018}
D.~Neuenfeld
[arXiv:1810.1477]


\bibitem{Carney:2018ygh}
D.~Carney, L.~Chaurette, D.~Neuenfeld and G.~Semenoff,
JHEP \textbf{09} (2018), 121
doi:10.1007/JHEP09(2018)121
[arXiv:1803.02370 [hep-th]].




\end{thebibliography}
\end{document}